\documentclass[11pt,aps,prd,nofootinbib,superscriptaddress,preprint,preprintnumbers
]{revtex4-2}
\usepackage{amssymb}
\usepackage{amsmath}
\usepackage{graphics}
\usepackage{epsfig}
\usepackage{multirow}
\usepackage{color}
\usepackage{mathrsfs}
\usepackage{graphicx}
\usepackage{slashed}
\usepackage{hyperref}

\begin{document}

\title{Scalar-pseudoscalar pair production at the Large Hadron Collider at NLO+NLL accuracy in QCD}

\author{He-Yi Li}
\affiliation{State Key Laboratory of Particle Detection and Electronics, University of Science and Technology of China, Hefei 230026, Anhui, People's Republic of China}
\affiliation{Department of Modern Physics, University of Science and Technology of China, Hefei 230026, Anhui, People's Republic of China}

\author{Ren-You Zhang}
\email{zhangry@ustc.edu.cn}
\affiliation{State Key Laboratory of Particle Detection and Electronics, University of Science and Technology of China, Hefei 230026, Anhui, People's Republic of China}
\affiliation{Department of Modern Physics, University of Science and Technology of China, Hefei 230026, Anhui, People's Republic of China}

\author{Yu Zhang}
\affiliation{Institutes of Physical Science and Information Technology, Anhui University, Hefei 230601, Anhui, People's Republic of China}
\affiliation{School of Physics and Materials Science, Anhui University, Hefei 230601, Anhui, People's Republic of China}

\author{Wen-Gan Ma}
\affiliation{State Key Laboratory of Particle Detection and Electronics, University of Science and Technology of China, Hefei 230026, Anhui, People's Republic of China}
\affiliation{Department of Modern Physics, University of Science and Technology of China, Hefei 230026, Anhui, People's Republic of China}

\author{Ming-Ming Long}
\affiliation{State Key Laboratory of Particle Detection and Electronics, University of Science and Technology of China, Hefei 230026, Anhui, People's Republic of China}
\affiliation{Department of Modern Physics, University of Science and Technology of China, Hefei 230026, Anhui, People's Republic of China}

\author{Shu-Xiang Li}
\affiliation{State Key Laboratory of Particle Detection and Electronics, University of Science and Technology of China, Hefei 230026, Anhui, People's Republic of China}
\affiliation{Department of Modern Physics, University of Science and Technology of China, Hefei 230026, Anhui, People's Republic of China}

\date{\today}

\begin{abstract}
We thoroughly investigate both transverse momentum and threshold resummation effects on scalar-pseudoscalar pair production via quark-antiquark annihilation at the $13~ \text{TeV}$ Large Hadron Collider at QCD NLO+NLL accuracy. A factorization method is introduced to properly supplement the soft-gluon (threshold) resummation contribution from parton distribution functions to the resummed results obtained by the  Collins-Soper-Sterman resummation approach. We find that the impact of the threshold-resummation improved PDFs is comparable to the resummation effect of the partonic matrix element and can even predominate in high invariant mass regions. Moreover, the loop-induced gluon-gluon fusion channel in the type-I two-Higgs-doublet model is considered in our calculation. The numerical results show that the electroweak production via quark-antiquark annihilation dominates over the gluon-initiated QCD production by $1 \sim 2$ orders of magnitude.
\end{abstract}

\maketitle

\section{Introduction}
\label{section-1}
\par
Primary tasks at the Large Hadron Collider (LHC) include the precision test of the standard model (SM) and the search for new physics beyond the SM (BSM). After the discovery of the $125~ \text{GeV}$ Higgs boson by both ATLAS and CMS collaborations at the LHC in 2012 \cite{Aad:2012tfa,Chatrchyan:2012xdj}, the SM has become the most successful theory in describing the interactions of fundamental particles. However, the discovery of this SM-like Higgs boson is merely one step toward fully investigating the electroweak symmetry breaking (EWSB). As is well known, the theoretical predictions of the SM are not always compatible with experimental observations, such as the dark matter in the universe, the oscillation of neutrinos, the huge hierarchy between electroweak and Planck scales, and the fine-tuning problem of Higgs mass. These conceptional and experimental difficulties encountered by the SM imply the existence of new physics beyond the SM.

\par
We may extend the SM by enlarging its gauge symmetry and/or introducing much more gauge multiplets to construct a new physics model. Among all the BSM theories, the two-Higgs-doublet model (2HDM) \cite{Branco:2011iw} is one of the simplest extensions of the SM. The Higgs sector responsible for the EWSB consists of two complex scalar isospin doublets, and the minimal supersymmetric standard model is a particular realization of the 2HDM. After EWSB, the three Goldstone modes $G^{\pm}$ and $G^0$ in the Higgs sector of the 2HDM are absorbed by the weak gauge bosons $W^{\pm}$ and $Z^0$, respectively, providing the longitudinal polarizations of $W^{\pm}$ and $Z^0$. The remaining five mass eigenstates of the Higgs sector are the so-called $\mathcal{CP}$-even Higgs bosons $h^0$ and $H^0$, $\mathcal{CP}$-odd Higgs boson $A^0$, and charged Higgs bosons $H^{\pm}$.

\par
Clearly, any discovery of a BSM Higgs boson will be an evidence for the existence of a new Higgs sector. At the LHC, the neutral Higgs bosons of the 2HDM can be produced both singly and in identical or mixed pairs. The dominant mechanism for single production of neutral Higgs bosons is gluon-gluon fusion. Concerning scalar-pseudoscalar pairs, the electroweak production via quark-antiquark annihilation, $pp \rightarrow q\bar{q} \rightarrow Z^{\ast} \rightarrow H^0A^0/h^0A^0$, can dominate over the QCD production via gluon-gluon fusion \cite{Hespel:2014sla, Enberg:2016ygw}, even by orders of magnitude. Considerable efforts have been devoted to search for BSM neutral Higgs bosons. Particularly, the exotic decays of heavy scalar (pseudoscalar), such as $H^0 \rightarrow A^0Z^0$ ($A^0 \rightarrow H^0Z^0$), have attracted attention at the LHC in recent years \cite{Sirunyan:2019wrn,Aad:2020ncx}. The scalar-pseudoscalar pair production is dominated by the Drell-Yan channel; it is an ideal process to investigate the Higgs gauge coupling $g_{H^0A^0Z^0}/g_{h^0A^0Z^0}$ and should thus be thoroughly investigated. The next-to-leading order (NLO) QCD corrections to neutral Higgs-boson pair production at hadron colliders were calculated in Refs.\cite{Dawson:1998py, Hespel:2014sla}, which showed that the QCD corrections can enhance the cross section of $h^0A^0$ production by approximately $30\%$.

\par
Fixed-order perturbative predictions would be unreliable when the exponential enhancement from soft gluon dominates at the edge of the phase space. The large logarithms, such as $\alpha_s^n ( M^2/p_T^2) \ln^m (M^2/p_T^2)$ at small-$p_T$ and $\alpha_s^n (1-z)^{-1} \ln^{m}(1-z)$ when $z = M^2/\hat{s} \rightarrow 1$, should be resummed in precision calculations. We extract such logarithms in the partonic matrix element by adopting the Collins-Soper-Sterman resummation technique \cite{Collins:1981va,Collins:1981uk,Collins:1984kg,Sterman:1986aj,Catani:1989ne,Catani:1996yz,Catani:2000vq,Bozzi:2005wk} and analyze the threshold resummation effect from parton distribution functions (PDFs) by using the factorization method proposed in Ref.\cite{Beenakker:2015rna}. Generally, the resummation corrections are only considered for fixing the unnatural behaviours in the small-$p_T$ and threshold regions, while the fixed-order predictions are suitable for describing the kinematics far away from the edge of the final-state phase space. Thus, the resummation results should be matched with the fixed-order predictions to obtain a reliable description in all kinematical regions.

\par
In this study, we thoroughly analyze scalar-pseudoscalar pair production at the $13~ \text{TeV}$ LHC within the type-I 2HDM at the NLO and next-to-leading logarithmic (NLL) accuracy in QCD. The rest of this paper is organized as follows. In Sec. \ref{section-2}, we briefly review the 2HDM. In Sec. \ref{section-3}, we present the calculation strategies for $\phi^0A^0$ associated production at QCD NLO+NLL accuracy, including the Collins-Soper-Sterman resummation technique and the factorization method for assessing the impact of the threshold-resummation improved PDFs. The numerical results and discussion for both the integrated cross section and the differential distributions with respect to the transverse momentum and invariant mass of the final-state $\phi^0A^0$ system are provided in Sec. \ref{section-4}. Finally, a short summary is given in Sec. \ref{section-5}.

\section{Brief review of 2HDM}
\label{section-2}
\par
In contrast to the SM, the Higgs sector of the 2HDM consists of two complex scalar $SU(2)_L$ doublets $\Phi_{1,2}$ with hypercharge $Y=+1$. The most general scalar potential, which is invariant under the $SU(2)_L \otimes U(1)_Y$ electroweak gauge symmetry and a discrete $Z_2$ symmetry $\Phi_i \rightarrow (-1)^{i+1} \Phi_i~ (i = 1, 2)$, is given by
\begin{equation}
\label{VHiggs}
\begin{aligned}
V(\Phi_{1}, \Phi_{2})
&=
m^2_{11} \Phi^{\dag}_{1} \Phi_{1}
+
m^2_{22} \Phi^{\dag}_{2} \Phi_{2}
-
( m^2_{12} \Phi^{\dag}_{1}\Phi_{2} + \text{h.c.} )
+
\frac{1}{2} \lambda_1 ( \Phi^{\dag}_{1} \Phi_{1} )^{2}
+
\frac{1}{2} \lambda_2 ( \Phi^{\dag}_{2} \Phi_{2} )^{2}  \\
&\quad
+
\lambda_{3} ( \Phi^{\dag}_{1} \Phi_{1} ) ( \Phi^{\dag}_{2} \Phi_{2} )
+
\lambda_{4} ( \Phi^{\dag}_{1} \Phi_{2} ) ( \Phi^{\dag}_{2} \Phi_{1} )
+
\dfrac{1}{2} \left[
\lambda_{5} (\Phi^{\dag}_{1}\Phi_{2})^{2}
+ \text{h.c.}
\right],
\end{aligned}
\end{equation}
where the dimension-two term $m_{12}^2$ is tolerated since it only breaks the $Z_2$ symmetry softly, and $m_{11, 22}^2$, $\lambda_{1, 2, 3, 4}$ are forced to be real due to the hermiticity of the scalar potential. The two Higgs doublets can be parameterized as \cite{Branco:2011iw}
\begin{equation}
    \Phi_{i}=
    \left(\begin{array}{c}{\phi_{i}^{+}} \\ {(v_{i}+\rho_{i}+i\eta_{i})/\sqrt{2} }\end{array}\right)
    \quad\quad
    (i=1,2),
\end{equation}
where $v_1$ and $v_2$ are the vacuum expectation values (VEVs) of the neutral components of $\Phi_{1}$ and $\Phi_{2}$, respectively. In a $\mathcal{CP}$-conserving 2HDM, both $m_{12}^2$ and $\lambda_5$ are real, and so are $v_1$ and $v_2$. The eight mass eigenstates of the Higgs sector are given by
\begin{equation}
\left(
      \begin{array}{c}
                H^0 \\
                h^0
      \end{array}
\right)
=
\left(
      \begin{array}{cc}
               \cos \alpha & \sin \alpha \\
              -\sin \alpha & \cos \alpha
      \end{array}
\right)
\left(
      \begin{array}{c}
               \rho_{1} \\
               \rho_{2}
      \end{array}
\right),
\end{equation}
\begin{equation}
\left(
      \begin{array}{c}
                G^0 \\
                A^0
      \end{array}
\right)
=
\left(
      \begin{array}{cc}
               \cos \beta & \sin \beta \\
              -\sin \beta  & \cos \beta
      \end{array}
\right)
\left(
      \begin{array}{c}
                \eta_{1} \\
                \eta_{2}
      \end{array}
\right),
\end{equation}
\begin{equation}
\left(
      \begin{array}{c}
               G^{\pm} \\
               H^{\pm}
      \end{array}
\right)
=
\left(
      \begin{array}{cc}
                \cos \beta & \sin \beta \\
               -\sin \beta  & \cos \beta
      \end{array}
\right)
\left(
      \begin{array}{c}
                \phi_{1}^{\pm} \\
                \phi_{2}^{\pm}
      \end{array}
\right),
\end{equation}
where $\alpha$ is the mixing angle in the $\mathcal{CP}$-even Higgs sector and $\beta = \arctan \dfrac{v_2}{v_1}$ describes the mixing in the $\mathcal{CP}$-odd and charged Higgs sectors. After the spontaneous electroweak symmetry breaking, three out of eight degrees of freedom from $\Phi_{1,2}$ that correspond to Nambu-Goldstone bosons $G^{\pm}$and $G^0$ are respectively absorbed by weak gauge bosons $W^{\pm}$ and $Z^0$, providing the longitudinal polarizations of $W^{\pm}$ and $Z^0$. The remaining five degrees of freedom become the aforementioned five physical Higgs bosons: two $\mathcal{CP}$-even Higgs bosons $h^0$ and $H^0$, one $\mathcal{CP}$-odd Higgs boson $A^0$, and a pair of charged Higgs bosons $H^{\pm}$. In this study, the seven input parameters for the Higgs sector of a $\mathcal{CP}$-conserving 2HDM are chosen as
\begin{equation}
\label{parfor2hdm}
\{ m_{h^0},\, m_{H^0},\, m_{A^0},\, m_{H^{\pm}},\, m^2_{12},\, \sin(\beta - \alpha),\, \tan\beta \},
\end{equation}
which are implemented as the ``physical basis'' in \textit{2HDMC} \cite{Eriksson:2010zzb}. Then, the Higgs potential in Eq.(\ref{VHiggs}) can be completely determined by above seven Higgs parameters in Eq.(\ref{parfor2hdm}) and $v$, where $v \equiv \sqrt{v_1^2+v_2^2} = (\sqrt{2}G_{F})^{-1/2} \approx 246 \, \text{GeV}$ has been classified as an electroweak input parameter.

\par
To guarantee the absence of Higgs-mediated flavor changing neutral currents at the tree level, the $Z_2$ symmetry should be extended to the Yukawa sector. Given that the two Higgs doublets $\Phi_{1,2}$ have opposite $Z_2$ charges, each flavor of quark/lepton can only couple to one of the two Higgs doublets. There are four allowed types of Yukawa interaction corresponding to the four independent $Z_2$ charge assignments on the quark and lepton $SU(2)_L$ multiplets (Table \ref{type-2HDM}). The Yukawa Lagrangian of the 2HDM can be expressed in terms of Higgs mass eigenstates as
\begin{equation}
\label{Lyukawa}
\begin{aligned}
\mathcal{L}_{\text{Yukawa}}^{\text{2HDM}}
=
\,
&- \sum_{f = u, d, \ell} \frac{m_f}{v}
\left(
\xi_h^f \bar{f} f h^0 + \xi_H^f \bar{f} f H^0 - i \xi_A^f \bar{f} \gamma^5 f A^0
\right) \\
&
-
\left[
\frac{\sqrt{2} V_{ud}}{v}
\bar{u}
\left(
m_u \xi_A^u P_L + m_d \xi_A^d P_R
\right)
d H^+
+
\frac{\sqrt{2} m_{\ell}}{v}
\xi_A^{\ell} \bar{\nu} P_R \ell H^+
+ \text{h.c.}
\right],
\end{aligned}
\end{equation}
where $\xi_{h, H, A}^{f}~ (f = u, d, \ell)$ are the Higgs Yukawa couplings normalized to the SM vertices, and the corresponding values in the type-I, type-II, lepton-specific, and flipped 2HDMs are listed in Table \ref{Mfactor}.
%-----------------------------------------------------Table 1-----------------------------------------------------
\begin{table}[!htbp]
\renewcommand \tabcolsep{10.0pt}
\centering
\begin{tabular}{|c|ccccccc|}
\hline\hline
2HDM & $\Phi_1$ & $\Phi_2$ & $Q_L$ & $L_L$ & $u_R$ & $d_R$ & $\ell_R$ \\
\hline
Type I   & $+$ & $-$ & $+$ & $+$ & $-$ & $-$ & $-$ \\
Type II  & $+$ & $-$ & $+$ & $+$ & $-$ & $+$ & $+$ \\
Lepton-specific & $+$ & $-$ & $+$ & $+$ & $-$ & $-$ & $+$ \\
Flipped  & $+$ & $-$ & $+$ & $+$ & $-$ & $+$ & $-$ \\
\hline\hline
\end{tabular}
\caption{\label{type-2HDM} Four types of 2HDMs and the corresponding $Z_2$ charge assignments on Higgs, quark, and lepton $SU(2)_L$ multiplets.}
\end{table}
%-----------------------------------------------------Table 2-----------------------------------------------------
\begin{table}[!htbp]
\renewcommand \tabcolsep{10.0pt}
\centering
\begin{tabular}{|c|c|c|c|c|}
\hline\hline
2HDM & Type I & Type II & Lepton-specific & Flipped \\
\hline
$\xi_h^u$
                              & $\quad\cos\alpha/\sin\beta\quad$
                              & \multirow{3}*{$\surd$}
                              & \multirow{6}*{$\surd$}
                              & \multirow{3}*{$\surd$} \\
$\xi_H^u$
                              & $\sin\alpha/\sin\beta$
                              &
                              &
                              & \\
$\xi_A^u$
                              & $\cot\beta$
                              &
                              &
                              & \\
\cline{3-3}\cline{5-5}
$\xi_h^d$
                              & $\cos\alpha/\sin\beta$
                              & \multirow{6}*{$(\alpha, \beta) \rightarrow (\tilde{\alpha}, \tilde{\beta})$}                              &
                              & \multirow{3}*{$(\alpha, \beta) \rightarrow (\tilde{\alpha}, \tilde{\beta})$} \\
$\xi_H^d$
                              & $\sin\alpha/\sin\beta$
                              &
                              &
                              & \\
$\xi_A^d$
                              & $-\cot\beta\quad$
                              &
                              &
                              & \\
\cline{4-5}
$\xi_h^{\ell}$
                              & $\cos\alpha/\sin\beta$
                              &
                              & \multirow{3}*{$(\alpha, \beta) \rightarrow (\tilde{\alpha}, \tilde{\beta})$}
                              & \multirow{3}*{$\surd$} \\
$\xi_H^{\ell}$
                              & $\sin\alpha/\sin\beta$
                              &
                              &
                              & \\
$\xi_A^{\ell}$
                              & $-\cot\beta\quad$
                              &
                              &
                              & \\
\hline\hline
\end{tabular}
\caption{\label{Mfactor} Normalized Higgs Yukawa couplings $\xi_{h, H, A}^{f}~ (f = u, d, \ell)$ in the type-I, type-II, lepton-specific, and flipped 2HDMs. $(\tilde{\alpha}, \tilde{\beta}) = (\alpha, \beta) + \dfrac{\pi}{2}$.}
\end{table}
%-------------------------------------------------------------------------------------------------------------------

\par
The Higgs gauge interaction is independent of the types of the 2HDM. The couplings of $h^0$ and $H^0$ to weak gauge boson pairs are proportional to $\sin(\beta - \alpha)$ and $\cos (\beta - \alpha)$, respectively. The 2HDM parameter space is stringently constrained by the requirement that one out of the two neutral $\mathcal{CP}$-even Higgs bosons has physical properties consistent with the $125~ \text{GeV}$ scalar discovered at the CERN LHC. It is well known that if one of the two neutral $\mathcal{CP}$-even Higgs mass eigenstates is approximately aligned in the two-dimensional Higgs field space with the direction of the Higgs VEV vector $\vec{v} \equiv (v_1, v_2)$ (the so-called alignment limit), the couplings of this Higgs boson are SM-like. The two alignment limits of the 2HDM are listed in Table \ref{A-limit}. Given that the SM-like Higgs boson with mass around $125~ \text{GeV}$ seems to be favored by LHC data, we will investigate the scalar-pseudoscalar pair production at the LHC only at the alignment limit.
%-----------------------------------------------------Table 3-----------------------------------------------------
\begin{table}[!htbp]
\renewcommand \tabcolsep{10.0pt}
\centering
\begin{tabular}{|c|ccc|}
\hline\hline
Alignment limit & $\beta - \alpha$ & $h^0$ & $H^0$ \\
\hline
I  & $\pi/2$ & SM-like & \\
II & $0$ &  & SM-like \\
\hline\hline
\end{tabular}
\caption{\label{A-limit} Two alignment limits of the 2HDM.}
\end{table}
%--------------------------------------------------------------------------------------------------------------------

\section{Calculation strategy}
\label{section-3}
\par
We adopt the ’t Hooft-Feynman gauge and take five-flavor scheme in our calculations. Apart from the top quark, all other light quarks, including the bottom quark, are treated as massless particles. The UV and IR divergences in the QCD loop and real jet emission corrections are regularized by adopting the dimensional regularization scheme \cite{tHooft:1972tcz}. We employ both the Catani-Seymour dipole subtraction method \cite{Catani:1996vz} and the two cutoff phase space slicing method \cite{Harris:2001sx} to separate the soft and collinear IR singularities of the real emission correction, and then cross-check their correctness.

\subsection{Electroweak production via quark-antiquark annihilation}
\par
\begin{figure}[htb]
\centering
\includegraphics[scale=0.45]{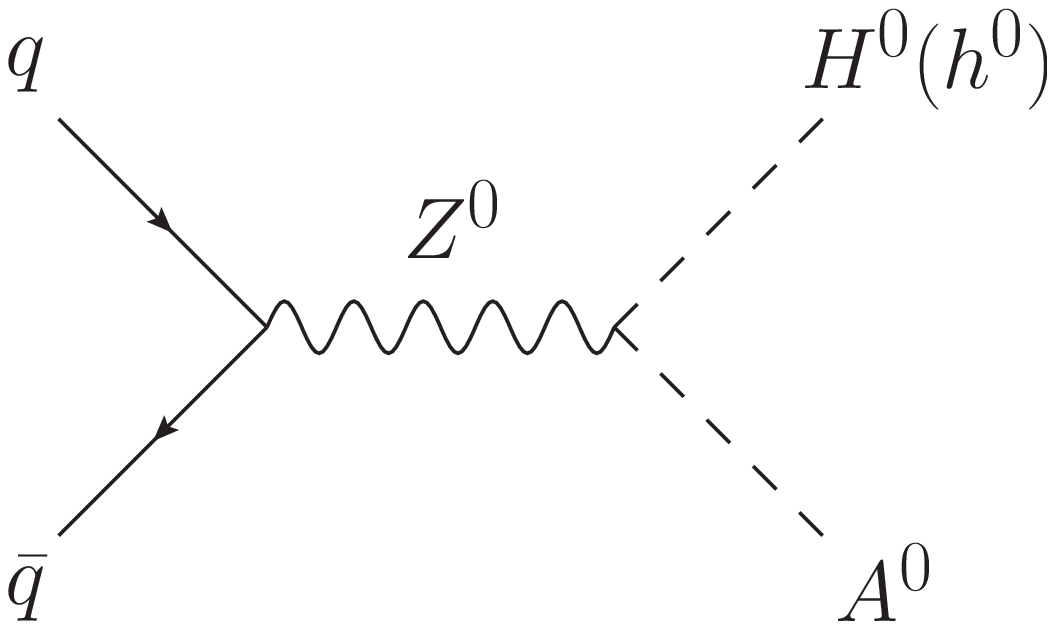}
\includegraphics[scale=0.45]{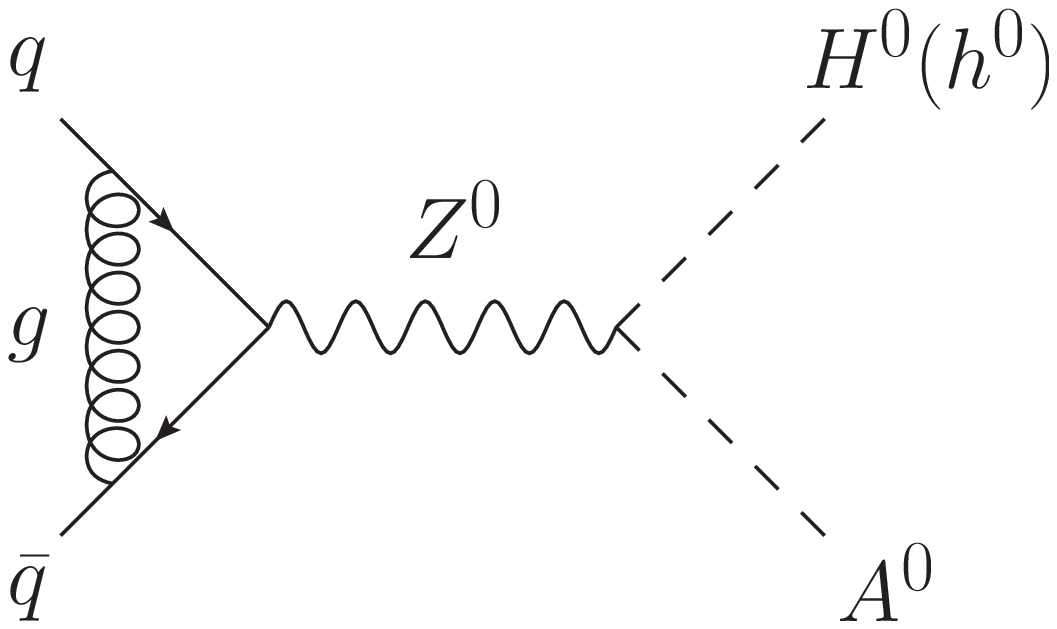}
\includegraphics[scale=0.45]{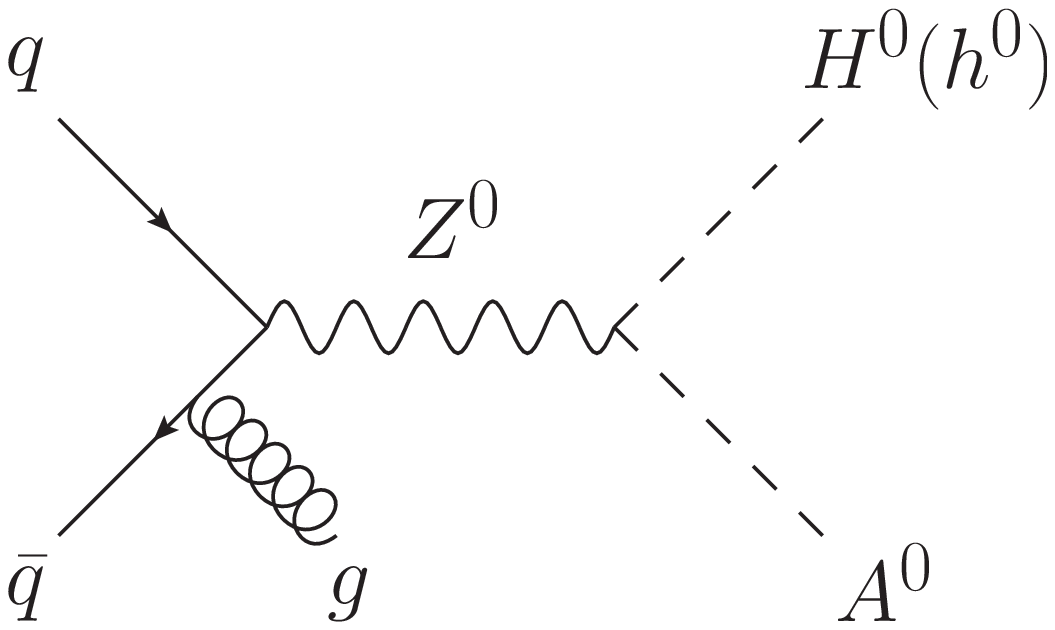}
\includegraphics[scale=0.45]{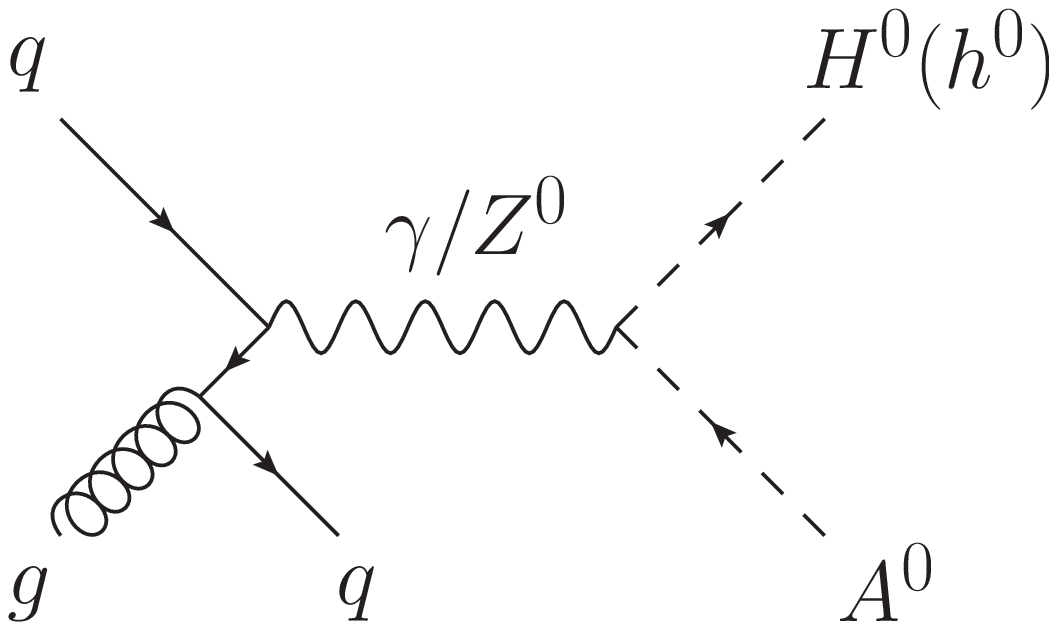}
\includegraphics[scale=0.45]{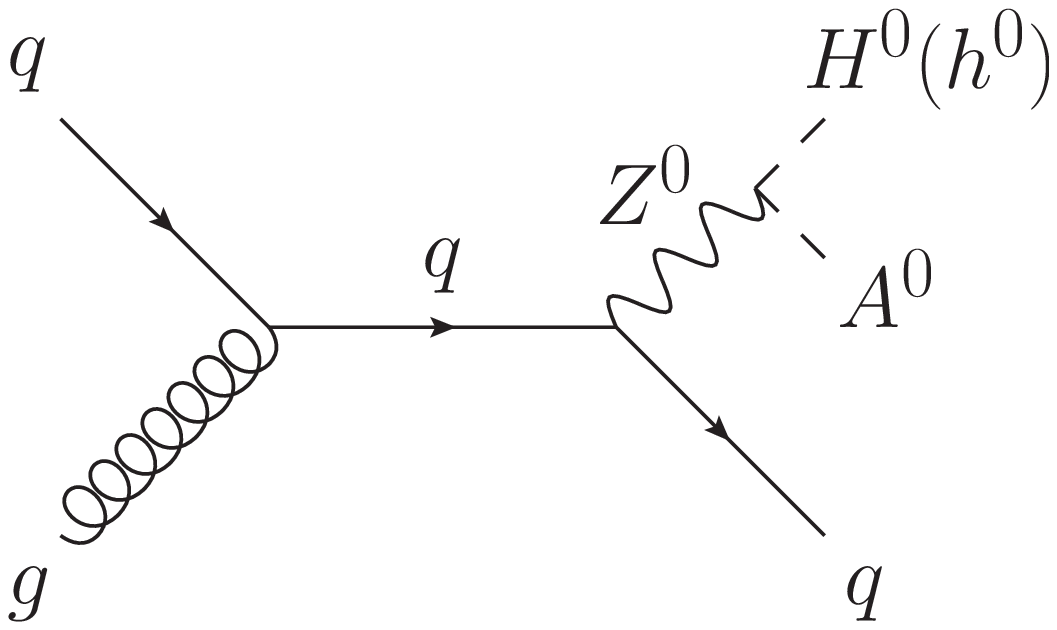}
\centering
\caption{Representative LO and QCD NLO Feynman diagrams for $q\bar{q} \rightarrow H^0(h^0)A^0$.}
\label{fig1}
\end{figure}

The scalar-pseudoscalar pair can be produced at the LHC via Drell-Yan production mechanism. Some representative LO and QCD NLO Feynman diagrams for $q \bar{q} \rightarrow H^0(h^0)A^0$ are shown in Fig.\ref{fig1}. In this study, we categorize $q g \rightarrow H^0(h^0)A^0 + q$ as the real light-quark emission correction to the quark-antiquark-initiated Drell-Yan channel.

\par
The $h^0A^0Z^0$ and $H^0A^0Z^0$ gauge interactions in the 2HDM are given by
\begin{equation}
\label{gspz}
g_{h^0A^0Z^0} = \frac{e \cos(\beta - \alpha)}{\sin2\theta_{\text{W}}} \left( p_{h^0}-p_{A^0} \right)_\mu,
\qquad
g_{H^0A^0Z^0} = -\frac{e \sin(\beta - \alpha)}{\sin2\theta_{\text{W}}} \left( p_{H^0}-p_{A^0} \right)_\mu,
\end{equation}
where $\theta_{\text{W}}$ is the Weinberg weak mixing angle, $p_{h^0, H^0, A^0}$ are the incoming momenta of the corresponding (pseudo)scalars, and $\mu$ is the Lorentz index of the vector boson $Z^0$. At the alignment limit, one of the two $\mathcal{CP}$-even mass eigenstates can be regarded as the SM Higgs boson $h^0_{\text{SM}}$, while the other is a BSM $\mathcal{CP}$-even Higgs boson denoted by $\phi^0$. We can see from Table \ref{A-limit} that
\begin{equation}
(h^0_{\text{SM}}\, , \, \phi^0)
=
\left\lbrace
\begin{aligned}
&~ (h^0, \, H^0)\,,
&\quad&
(\text{alignment limit I:} &&\sin(\beta - \alpha) = 1)
\\
&~ (H^0, \, h^0)\,,
&\quad&
(\text{alignment limit II:} &&\cos(\beta - \alpha) = 1)
\end{aligned}
\right.
\end{equation}
Given that the $h^0A^0Z^0$ and $H^0A^0Z^0$ coupling strengths are proportional to $\cos(\beta - \alpha)$ and $\sin(\beta - \alpha)$, respectively, the $h^0_{\text{SM}}A^0$ associated production is forbidden up to $\mathcal{O}(\alpha^2\alpha_s)$ at the alignment limit. Thus, we only focus on the Drell-Yan production of $\phi^0A^0$ in the following.
%-----------------------------------------------------Figure 1-----------------------------------------------------
%---------------------------------------------------------------------------------------------------------------------

\par
The doubly-differential cross section for $pp \rightarrow \phi^0A^0 + X$ can be perturbatively calculated by means of the QCD factorization theorem:
\begin{equation}
\label{Ftheorem}
\begin{aligned}
M^2 \frac{d^2\sigma}{dM^2 dp_T^2}(\tau)
=
&\sum_{a,b}
\int_0^1 dx_a dx_b dz
\Big[ x_a f_{a/P}(x_a, \mu_F^2) \Big]
\Big[ x_b f_{b/P}(x_b, \mu_F^2) \Big]
\\
&
\times \Big[ z \hat{\sigma}_{ab}(z, M^2, p_T^2, \mu_F^2, \mu_R^2) \Big]
\delta \big( \tau - x_a x_b z \big),
\end{aligned}
\end{equation}
where $M$ and $p_T$ are the invariant mass and transverse momentum of the final-state $\phi^0A^0$ system, respectively. The threshold variables $\tau$ and $z$ in Eq.(\ref{Ftheorem}) are defined by
\begin{equation}
\tau = \big( M/\sqrt{s} \big)^2,
\qquad\qquad
z = \big( M/\sqrt{\hat{s}} \big)^2,
\end{equation}
where $\sqrt{s}$ and $\sqrt{\hat{s}}$ denote the hadronic and partonic center-of-mass energies, respectively. The universal PDF $f_{a/P}(x, \mu_F^2)$ gives the probability to find parton $a$ in proton $P$ at factorization scale $\mu_F$ as a function of fraction $x$ of the proton's longitudinal momentum carried by the parton. After preforming a Mellin transformation,
\begin{equation}
F(N)= \int_0^1 dy y^{N-1} F(y),
\end{equation}
on Eq.(\ref{Ftheorem}), the hadronic cross section can be written as a simple product of the PDFs and the partonic cross section in the conjugate Mellin $N$-space as
\begin{equation}
\label{fac-N}
M^2 \frac{d^2\sigma}{dM^2dp_T^2}(N-1)
=
\sum_{a,b} f_{a/P}(N, \mu_F^2) f_{b/P}(N, \mu_F^2) \hat{\sigma}_{ab}(N, M^2, p_T^2, \mu_F^2, \mu_R^2).
\end{equation}
To be consistent with the CT collaboration \cite{Dulat:2015mca}, we refit the PDF, $f_{a/P}(x, \mu_F^2)$, as a polynomial of $x^{1/2}$ with eight coefficients,
\begin{equation}
f_{a/P}(x, \mu_F^2)
=
A_0 x^{A_1}
\left( 1 - x \right)^{A_2}
\left( 1 + A_3 x^{1/2} + A_4 x + A_5 x^{3/2} + A_6 x^{2} + A_7 x^{5/2} \right).
\end{equation}
Thus, the Mellin moment of the PDF has the form
\begin{equation}
\begin{aligned}
f_{a/P}(N, \mu_F)
&=
A_0
\Big[
\text{B}(A_1+N, A_2+1)
+ A_3 \text{B} (A_1+N+1/2, A_2+1) \\
& \quad
+ A_4 \text{B} (A_1+N+1, A_2+1)
+ A_5 \text{B} (A_1+N+3/2, A_2+1) \\
& \quad
+ A_6 \text{B} (A_1+N+2, A_2+1)
+ A_7 \text{B} (A_1+N+5/2, A_2+1)
\Big],
\end{aligned}
\end{equation}
where $\text{B}(x, y) \equiv \Gamma(x)\Gamma(y)/\Gamma(x+y)$ is the \textit{Beta} function.

\par
According to the factorization scheme presented in Ref.\cite{Contopanagos:1996nh}, the partonic cross section can be expressed as a product of a process-dependent hard function and a process-independent Sudakov exponential term \cite{Sterman:1986aj,Catani:1989ne,Catani:1990rp,Kidonakis:1997gm,Kidonakis:1998bk,Vogt:2000ci}. The higher-order QCD contributions to the partonic cross section contain logarithmic terms of type $\alpha_s^n (M^2/p_T^2) \ln^m (M^2/p_T^2)$, which become large in the small-$p_T$ region. These logarithmically-enhanced contributions arising at small $p_T$ spoil the convergence of the fixed-order perturbative expansion and must therefore be resummed to all orders in $\alpha_s$. We adopt the $b$-space resummation approach, which was fully formulated by Collins, Soper and Sterman \cite{Collins:1981va,Collins:1981uk,Collins:1984kg}, to systematically resum the large logarithmic terms at small $p_T$. In this approach, a Bessel transform is applied to the partonic cross section,
\begin{equation}
\label{Bessel-transform}
\hat{\sigma}_{ab}(N, M^2, p_T^2, \mu_F^2, \mu_R^2)
=
\int_0^{\infty} db \frac{b}{2} J_0(b p_T)
\hat{\sigma}_{ab}(N, M^2, b^2, \mu_F^2, \mu_R^2),
\end{equation}
where $J_0(x)$ is the zeroth-order Bessel function. Given that impact parameter $b$ is the variable conjugate to transverse momentum $p_T$, the limit $M/p_T \rightarrow 0$ corresponds to $M b \rightarrow \infty$. Therefore, the large logarithms of $M/p_T$ arising at small $p_T$ turn into large logarithms of $Mb$,
\begin{equation}
\Big[ (M^2/p_T^2) \ln^m (M^2/p_T^2) \Big]_+
\longrightarrow
\ln^{m+1} (M^2b^2)
+ \cdots
\end{equation}
After performing the resummation procedure, the resummed partonic cross section in the conjugate $b$-space at the NLL accuracy can be expressed as \cite{Collins:1981va,Collins:1981uk,Collins:1984kg}
\begin{equation}
\label{res-pT}
\begin{aligned}
&
\hat{\sigma}_{ab}^{\text{(res.)}}(N, M^2, b^2, \mu_F^2, \mu_R^2)
=
\sum_{a^{\prime}, a^{\prime\prime}, b^{\prime}, b^{\prime\prime}}
E_{a^{\prime} a}^{(1)}(N, 1/\bar{b}^2, \mu_F^2)
E_{b^{\prime} b}^{(1)}(N, 1/\bar{b}^2, \mu_F^2) \\
&
\qquad\qquad
\times
\mathcal{C}_{a^{\prime\prime} a^{\prime}}(N, 1/\bar{b}^2)
\mathcal{C}_{b^{\prime\prime} b^{\prime}}(N, 1/\bar{b}^2)
\mathcal{H}_{a^{\prime\prime} b^{\prime\prime}}(M^2, \mu_R^2)
\exp
\Big[
\mathcal{G}_{a^{\prime\prime} b^{\prime\prime}}(M^2\bar{b}^2, M^2, \mu_R^2)
\Big],
\end{aligned}
\end{equation}
where $\bar{b}$ is the normalized impact parameter defined by $\bar{b} = b/b_0$ with $b_0 = 2 e^{-\gamma_{\text{E}}}$ \cite{Bozzi:2005wk}, and the one-loop QCD evolution operator $E^{(1)}_{ab}$ is derived from the collinear-improved procedure as recommended in Refs.\cite{Kramer:1996iq,Catani:2001ic,Kulesza:2002rh,Bozzi:2007qr,Almeida:2009jt}. In the physical resummation scheme, the coefficient function $\mathcal{C}_{ab}$ and the Sudakov form factor $\mathcal{G}_{ab}$ are free from any hard contributions, and the hard function $\mathcal{H}_{ab}$, determined by the finite part of the renormalized virtual contribution, is free from any logarithmic contributions \cite{Catani:2000vq}. To transform the resummed partonic cross section $\hat{\sigma}_{ab}^{\text{(res.)}}(N, M^2, b^2, \mu_F^2, \mu_R^2)$ back to the physical $p_T$-space, we rewrite Eq.(\ref{Bessel-transform}) as \cite{Laenen:2000de}
\begin{equation}
\hat{\sigma}_{ab}^{\text{(res.)}}(N, M^2, p_T^2, \mu_F^2, \mu_R^2)
=
\sum_{k=1,2}\,
\intop_{C_k} db \frac{b}{4}
h_k(b p_T, v)
\hat{\sigma}_{ab}^{\text{(res.)}}(N, M^2, b^2, \mu_F^2, \mu_R^2),
\end{equation}
where
\begin{equation}
\qquad
h_k(x, v) = \frac{(-1)^k}{\pi} \int_{-iv\pi}^{(-1)^k \pi + iv\pi}
d\theta e^{-i x \sin\theta}
\qquad
(k = 1, 2)
\end{equation}
are two auxiliary Hankel-like functions satisfying $h_1(x, v) + h_2(x, v) = 2 J_0(x)$, and the integration contours $C_k~ (k = 1, 2)$ in the complex $b$-plane are defined by
\begin{equation}
\quad
C_k:
\quad
b = b(t) \equiv t e^{(-1)^k i \varphi},
\qquad
t \in [0, +\infty)
\quad\text{with}\quad
\varphi \in (0, \pi/2).
\end{equation}
It is well known that such contours avoid the Landau pole by a deformation into either the upper or lower half complex $b$-plane.

\par
The invariant mass distribution of the final-state $\phi^0A^0$ system in the Mellin $N$-space can be obtained by integrating Eq.(\ref{fac-N}) over the transverse momentum $p_T$. In the threshold regime, the large logarithmic terms of the type $\alpha_s^n (1-z)^{-1} \ln^m (1-z)$ also spoil the convergence of the perturbative series. These singular terms turn into large logarithms of the Mellin variable, $N$:
\begin{equation}
\Big[ (1-z)^{-1} \ln^m (1-z) \Big]_+
\longrightarrow
\ln^{m+1} N
+ \cdots
\end{equation}
The corresponding resummed partonic cross section for invariant mass distribution at the NLL accuracy can be expressed as \cite{Furmanski:1981cw,Debove:2010kf}
\begin{equation}
\label{res-M}
\begin{aligned}
\hat{\sigma}_{ab}^{\text{(res.)}}(N, M^2, \mu_F^2, \mu_R^2)
&=
\sum_{a^{\prime}, b^{\prime}}
E_{a^{\prime} a}^{(1)}(N, M^2/\bar{N}^2, \mu_F^2)
E_{b^{\prime} b}^{(1)}(N, M^2/\bar{N}^2, \mu_F^2) \\
&
\quad
\times
\tilde{\mathcal{H}}_{a^{\prime} b^{\prime}}(M^2, \mu_R^2)
\exp
\Big[
\tilde{\mathcal{G}}_{a^{\prime} b^{\prime}}(\bar{N}, M^2, \mu_R^2)
\Big],
\end{aligned}
\end{equation}
where $\bar{N}$ is the reduced Mellin variable defined by $\bar{N} = N e^{\gamma_{\text{E}}}$.

\par
The hard functions, $\mathcal{H}_{ab}$ and $\tilde{\mathcal{H}}_{ab}$, do not contain any large logarithms. They can be perturbatively calculated and read at the NLO accuracy:
\begin{equation}
\begin{aligned}
\mathcal{H}_{ab}(M^2, \mu_R^2)
&=
\hat{\sigma}^{(0)}_{ab}(M^2)
\left( 1+a_s \mathcal{A}_0 \right),
\\
\tilde{\mathcal{H}}_{ab}(M^2, \mu_R^2)
&=
\mathcal{H}_{ab}(M^2, \mu_R^2)
+
a_s
\frac{\pi^2}{6}
\Big[
A_a^{(1)} + A_b^{(1)}
\Big]
\hat{\sigma}^{(0)}_{ab}(M^2),
\end{aligned}
\end{equation}
where $a_s = \alpha_s/(2\pi)$, $A_a^{(1)} = 2 C_a$\footnote{$C_q = C_F = 4/3$ and $C_g = C_A = 3$.}, $\hat{\sigma}_{ab}^{(0)}$ is the lowest-order partonic cross section, and $\mathcal{A}_0$ represents the IR-finite part of the renormalized virtual correction in the dimensional regularization scheme, i.e.,
\begin{equation}
\hat{\sigma}_{ab}^{\text{(vir.)}}(M^2, \mu_R^2)
=
a_s
\left( \frac{4\pi\mu_R^2}{M^2} \right)^{\epsilon} \frac{\Gamma(1 - \epsilon)}{\Gamma(1 - 2 \epsilon)}
\left(
\frac{\mathcal{A}_{-2}}{\epsilon^2} + \frac{\mathcal{A}_{-1}}{\epsilon} + \mathcal{A}_{0}
\right)
\hat{\sigma}_{ab}^{(0)}(M^2)
+
\mathcal{O}(\epsilon).
\end{equation}
The Sudakov form factors $\mathcal{G}_{ab}$ and $\tilde{\mathcal{G}}_{ab}$ collect all the logarithmically-enhanced contributions and take the form
\begin{equation}
\label{s-factor}
\mathbb{G}_{ab}(\omega, M^2, \mu_R^2)
=
L \mathbb{G}_{ab}^{(1)}(\lambda)
+
\sum_{n=0}^{+\infty}
a_s^{n} \mathbb{G}_{ab}^{(n+2)}(\lambda, M^2/\mu_R^2)
\qquad
(\mathbb{G} = \mathcal{G} \text{~or~} \tilde{\mathcal{G}}),
\end{equation}
with $\lambda = a_s \beta_0 L$, $L = \ln \omega$, and $\omega = M^2\bar{b}^2$ and $\bar{N}$ for $\mathbb{G} = \mathcal{G}$ and $\tilde{\mathcal{G}}$, respectively. The function $\mathbb{G}_{ab}^{(n+1)}~ (n=0,1,2,...)$ on the right side of Eq.(\ref{s-factor}) resums all the $\text{N}^{n}\text{LL}$ contributions. In this study, we only consider the LL and NLL terms, i.e., $L \mathbb{G}_{ab}^{(1)}$ and $\mathbb{G}_{ab}^{(2)}$, since the electroweak production of $\phi^0A^0$ is studied at the NLO+NLL accuracy. The analytic expressions for $\mathbb{G}_{ab}^{(1)}$ and $\mathbb{G}_{ab}^{(2)}$ can be found in Ref.\cite{Fuks:2013vua}. Finally, the $\mathcal{C}_{ab}$ function in Eq.(\ref{res-pT}) at the NLL accuracy can be expressed as
\begin{equation}
\mathcal{C}_{ab}(N, \mu_R^2)
=
\delta_{ab}
+
a_s
\left[
\frac{\pi^2}{6} C_{a} \delta_{ab} - P^{\prime}_{ab}(N)
\right],
\end{equation}
where $P_{ab}^{\prime}(N)$ is the $\mathcal{O}(\epsilon)$ part of the unregulated Altarelli-Parisi splitting function in the Mellin $N$-space, i.e.,
\begin{equation}
P_{ab}(z, \epsilon)
=
P_{ab}(z) + \epsilon P_{ab}^{\prime}(z)
\qquad\text{and}\qquad
P^{\prime}_{ab}(N)
=
\int_0^1 dz z^{N-1} P^{\prime}_{ab}(z).
\end{equation}

\par
The resummed partonic cross section $\hat{\sigma}_{ab}^{\text{(res.)}}$ gives the dominant contribution in the small-$p_T$ and threshold regions, while the fixed-order partonic cross section $\hat{\sigma}_{ab}^{\text{(f.o.)}}$ dominates at large $p_T$ and small $M/\sqrt{\hat{s}}$. To obtain a reliable theoretical prediction with uniform accuracy in all kinematical regions, the resummed and fixed-order results should be combined consistently by subtracting their overlap,
\begin{equation}
\label{combined-x}
\hat{\sigma}_{ab}
=
\hat{\sigma}_{ab}^{\text{(res.)}}
+
\hat{\sigma}_{ab}^{\text{(f.o.)}}
-
\hat{\sigma}_{ab}^{\text{(o.l.)}}.
\end{equation}
This matching procedure guarantees that the combined result $\hat{\sigma}_{ab}$ contains both the perturbative contributions up to the specific fixed order and the logarithmically-enhanced contributions from higher orders. At the NLO+NLL accuracy, $\hat{\sigma}_{ab}^{\text{(o.l.)}}$ in Eq.(\ref{combined-x}) can be obtained by expanding the resummed partonic cross section $\hat{\sigma}_{ab}^{\text{(res.)}}$ to $\mathcal{O}(\alpha_s)$, i.e.,
\begin{equation}
\hat{\sigma}_{ab}^{\text{(o.l.)}}
=
\hat{\sigma}_{ab}^{\text{(res.)}}
(\alpha_s=0)
+
\alpha_s
\frac{d \hat{\sigma}_{ab}^{\text{(res.)}}}{d \alpha_s}
(\alpha_s=0).
\end{equation}
After multiplying the Mellin moments of the PDFs to the NLO+NLL matched partonic cross section $\hat{\sigma}_{ab}$, we obtain the hadronic differential cross section in the Mellin $N$-space. To get back to the physical space, an inverse Mellin transform,
\begin{equation}
F(\tau)
=
\frac{1}{2 \pi i} \int_{C_N} dN \tau^{-N} F(N),
\end{equation}
should be applied to the right side of Eq.(\ref{fac-N}). To achieve this, we must comprehensively estimate the singularities in the Mellin $N$-space and choose an appropriate integration contour $C_N$. There are two types of singularities for the hadronic differential cross section in the Mellin $N$-space: (1) the poles in the Mellin moments of the PDFs (Regge poles), and (2) the Landau pole related to the running of the strong coupling constant. The integration contour $C_N$ in the complex $N$-plane is chosen as \cite{Catani:1996yz}
\begin{equation}
C_N:
\quad
N = N(y) \equiv C + y e^{\pm i \phi},
\qquad
y \in [0, +\infty)
\end{equation}
where $\phi \in [\pi/2, \pi)$ and the constant $C$ is chosen such that the Regge and Landau poles lie to the left and right of $C_N$, respectively.

\par
In principle, for NLO+NLL calculations, we should employ resummation-improved PDFs for initial-state parton convolution. The threshold-resummation improved PDFs are now available with the NNPDF3.0 set. Compared to the NNPDF3.0 global fit, the threshold-resummation improved PDF fit has to be performed with a reduced data set involving deep-inelastic scattering, Drell-Yan, and top-pair production data, because the threshold resummation calculations are not readily available for all the processes employed in the global analysis. The reduced data set used in the fit of the threshold-resummation improved PDF set would induce a relatively larger PDF error compared to the global PDF set. In this study, we adopt the factorization method proposed in Ref.\cite{Beenakker:2015rna} to combine the smaller PDF error of the global PDF set with the resummation effect from the threshold-resummation improved PDF set. In the factorization method, the NLO+NLL QCD corrected cross section can be approximately calculated by
\begin{equation}
\label{K-method}
\sigma^{\text{NLO+NLL}}
=
K
\times
\sigma^{\text{NLO}}\big|_{\text{(\textit{NLO global})}}\,,
\end{equation}
where
\begin{equation}
K = K_{\text{PDF}} \times K_{\text{PME}}\,,
\end{equation}
and
\begin{equation}
\label{K-pdf-pme}
K_{\text{PDF}}
=
\dfrac{\sigma^{\text{NLO+NLL}}\big|_{\text{(\textit{NLO+NLL reduced})}}}{\sigma^{\text{NLO+NLL}}\big|_{\text{(\textit{NLO reduced})}}}\,,
\qquad\qquad
K_{\text{PME}}
=
\dfrac{\sigma^{\text{NLO+NLL}}\big|_{\text{(\textit{NLO global})}}}{\sigma^{\text{NLO}}\big|_{\text{(\textit{NLO global})}}}
\end{equation}
which describe the impact of the threshold-resummation improved PDFs and the NLL resummation effect from the partonic matrix element, respectively. Subscripts ``\textit{NLO+NLL reduced}" and ``\textit{NLO reduced}" in the definition of $K_{\text{PDF}}$ denote the threshold-resummation improved PDF set NNPDF30\_nll\_disdytop and its fixed-order version NNPDF30\_nlo\_disdytop \cite{Bonvini:2015ira}, respectively. It is well known that the NNPDF cannot be properly transformed to the Mellin space; the refit of the NNPDF replicas in Mellin space would lead to some convergence issues \cite{Fuks:2013lya}. Fortunately, however, $K_{\text{PME}}$ is expected to be largely independent of the PDF choice since the PDF sets used in $K_{\text{PME}}$ are estimated at the same perturbative order \cite{Fuks:2013lya}. This feature has been verified with the CT18NLO and MSTW2008nlo68cl PDFs, and thus we choose CT18NLO as the ``\textit{NLO global}" PDF set in our calculations.

\subsection{QCD production via gluon-gluon fusion}
\par
\begin{figure}[htb]
\centering
\includegraphics[scale=0.45]{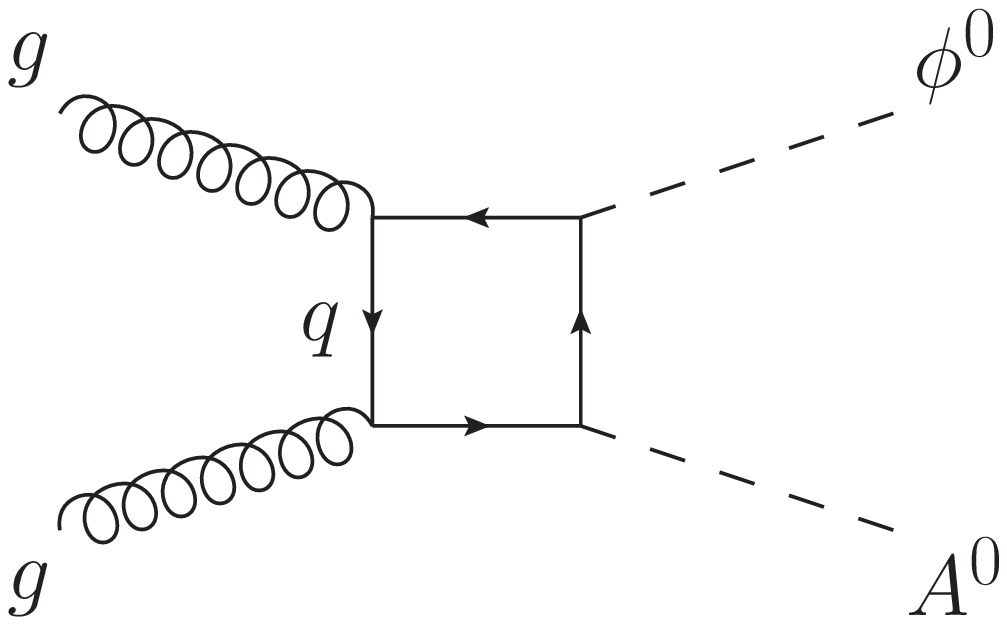}
\includegraphics[scale=0.45]{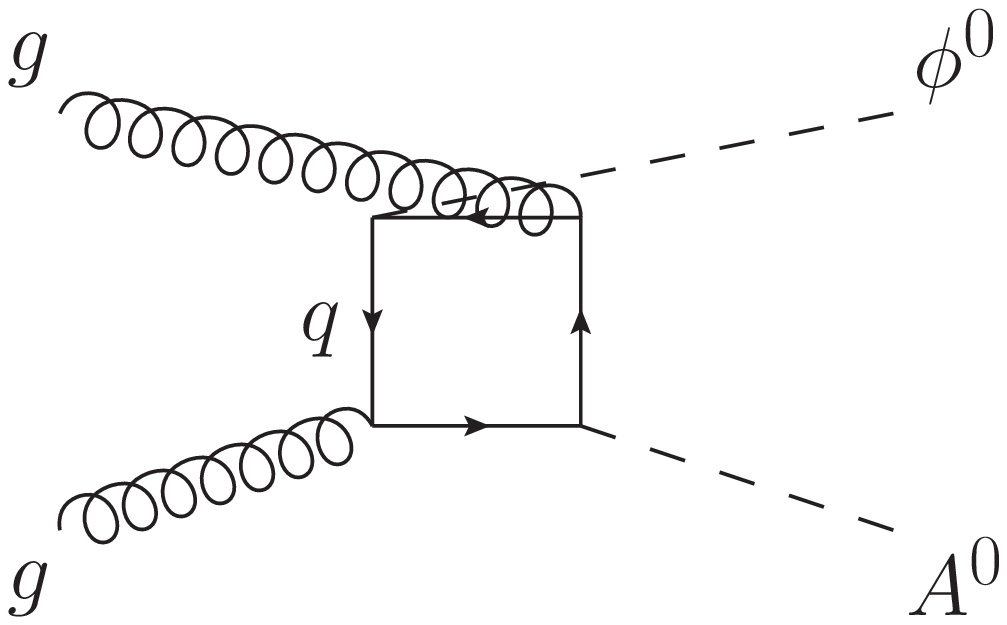}
\includegraphics[scale=0.45]{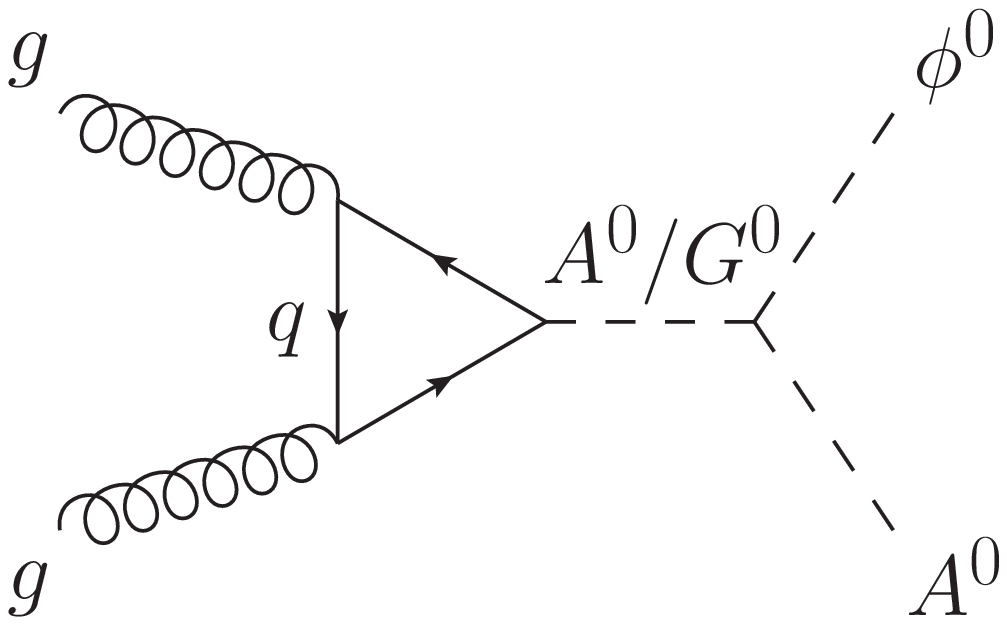}
\centering
\caption{Representative Feynman diagrams for $gg \rightarrow \phi^0A^0$ at the lowest order.}
\label{fig2}
\end{figure}

Compared to the electroweak production via quark-antiquark annihilation, the gluon-initiated QCD production of the scalar-pseudoscalar pair is a loop-induced production channel. This production mechanism should be taken into consideration at the LHC due to the high luminosity of gluon in proton. In Fig.\ref{fig2} we depict some representative Feynman diagrams for $gg \rightarrow \phi^0A^0$ at the lowest order. Note that the production rate relies not only on the heavy-quark Yukawa couplings, but also on the triple Higgs self-couplings. Unlike the quark-antiquark annihilation channel, the loop-induced gluon-gluon fusion channel is extremely sensitive to the Yukwawa interaction of the 2HDM. Due to the introduction of a soft breaking $Z_2$ symmetry to avoid tree-level FCNCs, each fermion type is only able to couple to one of the two Higgs doublets. There are four allowed types of 2HDMs, type-I, type-II, lepton-specific, and flipped, which correspond to the four different types of Yukawa interaction. In this study, we mainly focus on the type-I 2HDM and calculate the gluon-gluon fusion channel by using the modified \textit{FeynArts-3.9}, \textit{FormCalc-7.3}, and \textit{LoopTools-2.8} packages \cite{Hahn:1998yk, Hahn:2000kx, vanOldenborgh:1990yc}.

\section{Numerical results and discussion}
\label{section-4}
\par
In this section, we provide some numerical results for $pp \rightarrow \phi^0 A^0 + X$ at the $13~ \text{TeV}$ LHC  in the type-I 2HDM. The SM input parameters used in this study are taken as \cite{Zyla:2020zbs}
\begin{equation}
\begin{aligned}
&
m_W = 80.379 ~ \text{GeV},
\qquad~~
m_Z = 91.1876 ~ \text{GeV},
\qquad
&&
m_t = 172.76 ~ \text{GeV},
\\
&
G_{F} = 1.1663787 \times 10^{-5} ~ \text{GeV}^{-2},
&&
\alpha_s(m_Z) = 0.118.
\end{aligned}
\end{equation}
The input parameters for the Higgs sector of the 2HDM should satisfy the theoretical constraints from perturbative unitarity \cite{Grinstein:2015rtl}, stability of vacuum \cite{Nie:1998yn} and tree-level unitarity \cite{Akeroyd:2000wc}, which can be checked by \textit{2HDMC} \cite{Eriksson:2010zzb}. Moreover, the high-energy experiments can also give stringent constraints on those 2HDM input parameters. One of the experimental limits is that the partial width of $Z^0 \rightarrow H^0A^0/h^0A^0$ cannot exceed 2$\sigma$ uncertainty of the $Z$-width measurement \cite{Zyla:2020zbs}, and others come from the restriction of the physical observables of $B$ meson decays, the measurement of the SM-like Higgs property, and the direct search of Higgs state at the LEP, Tevatron, and LHC, which are integrated in the \textit{SuperIso} \cite{Mahmoudi:2008tp}, \textit{HiggsSignals} \cite{Bechtle:2013xfa}, and \textit{Higgsbounds} \cite{Bechtle:2020pkv} packages, respectively.

\par
We use the CT14lo PDF set \cite{Dulat:2015mca} to perform LO calculation, and employ the CT18NLO PDF \cite{Hou:2019qau} to obtain NLO and NLO+NLL QCD corrected cross sections. CT18NLO PDF contains $1$ central PDF set and $2 \times 29$ Hessian replicas. The PDF uncertainties of a cross section $\sigma$ calculated with the CT18NLO PDF are given by \cite{Pumplin:2001ct}
\begin{equation}
\begin{aligned}
\delta^+_{\text{PDF}}
&=
\frac{1}{\sigma_0}
\sqrt{
\sum_{i=1}^{29}
\Big[
\max \big\{ \sigma_{i+} - \sigma_0,\, \sigma_{i-} - \sigma_0,\, 0 \big\}
\Big]^2
}\,,
\\
\delta^-_{\text{PDF}}
&=
\frac{1}{\sigma_0}
\sqrt{
\sum_{i=1}^{29}
\Big[
\max \big\{ \sigma_0 - \sigma_{i+},\, \sigma_0 - \sigma_{i-},\, 0 \big\}
\Big]^2
}\,,
\end{aligned}
\end{equation}
where $\sigma_0$ is the central value calculated with central set and $\sigma_{i\pm}~ (i = 1, ..., 29)$ are the cross sections evaluated with replicas. The factorization scale $\mu_F$ and the renormalization scale $\mu_R$ are taken to be equal, i.e., $\mu_F = \mu_R = \mu$, for simplicity. The scale uncertainties of an integrated cross section $\sigma$ are defined by
\begin{equation}
\begin{aligned}
\delta^+_{\mu}
&=
\frac{\max \big\{ \sigma(\mu)\, \big| \, \mu_0/2 \leqslant \mu \leqslant 2\mu_0 \big\} - \sigma(\mu_0)}{\sigma(\mu_0)}\,,
\\
\delta^-_{\mu}
&=
\frac{\min \big\{ \sigma(\mu)\, \big| \, \mu_0/2 \leqslant \mu \leqslant 2\mu_0 \big\} - \sigma(\mu_0)}{\sigma(\mu_0)}\,,
\end{aligned}
\end{equation}
where $\mu_0$ is the central scale. The total theoretical error is defined as the sum in quadrature of the PDF and scale uncertainties. For the quark-initiated Drell-Yan production channel, $pp \rightarrow q\bar{q} \rightarrow \phi^0A^0$, the production rate will be calculated at the NLO+NLL accuracy in QCD, and the NLO and NLO+NLL relative corrections are respectively defined as
\begin{equation}
\delta^{\text{NLO}} = \frac{\sigma^{\text{NLO}} - \sigma^{\text{LO}}}{\sigma^{\text{LO}}},
\qquad\qquad
\delta^{\text{NLO+NLL}} = \frac{\sigma^{\text{NLO+NLL}} - \sigma^{\text{LO}}}{\sigma^{\text{LO}}}.
\end{equation}
Regarding the gluon-gluon fusion channel, $pp \rightarrow gg \rightarrow \phi^0A^0$, we only consider the lowest-order contribution since it is a loop-induced channel.

\subsection{Integrated cross section}
\par
In this subsection, we present the integrated cross sections for $\phi^0A^0$ associated production at $\sqrt{s} = 13~ \text{TeV}$ LHC at the alignment limit in the 2HDM. The masses of $\phi^0$ and $A^0$ can be alternatively described by the following three parameters,
\begin{equation}
m = \min\left( m_{\phi^0},\, m_{A^0} \right),
\qquad
\Delta m = \left| m_{\phi^0} - m_{A^0} \right|,
\qquad
\epsilon = \text{sign} \left( m_{\phi^0} - m_{A^0} \right),
\end{equation}
i.e., the minimal mass and mass hierarchy of $\phi^0$ and $A^0$. The two scenarios in which $\epsilon = +1$ ($m_{\phi^0} > m_{A^0}$) and $\epsilon = -1$ ($m_{\phi^0} < m_{A^0}$) may be referred to as the normal mass hierarchy and inverted mass hierarchy, respectively.

\par
The Drell-Yan production channel depends only on the masses of $\phi^0$ and $A^0$ \footnote{$\sin(\beta - \alpha)$ has been fixed at the alignment limit.}. Furthermore, its integrated cross section is independent of $\epsilon$. In our calculations, we take $0$, $50$, and $100~ \text{GeV}$ as three benchmark values of $\Delta m$, which correspond, respectively, to the following three $\phi^0$-$A^0$ mass splitting scenarios:
\begin{itemize}
\item degenerate scenario: $\Delta m = 0$
\item hierarchical scenario with small mass splitting: $0< \Delta m < m_Z$
\item hierarchical scenario with large mass splitting: $\Delta m > m_Z$
\end{itemize}
The LO, NLO, NLO+NLL QCD corrected integrated cross sections and the corresponding theoretical relative errors induced by the factorization/renormalization scale and PDFs for $pp \rightarrow q\bar{q} \rightarrow \phi^0A^0$ as functions of $m$ for $\Delta m = 0$, $50$, and $100~ \text{GeV}$ are given in Tables \ref{Xsection-0}, \ref{Xsection-50}, and \ref{Xsection-100}, respectively. The central scale is taken as $\mu_0 = m_{\phi^0} + m_{A^0}$. There is no PDF-induced theoretical error for the LO cross section, since the CT14lo PDF used in the LO calculation contains only one central set but no PDF replicas. To study the full NLL resummation effect and the impact of the threshold-resummation improved PDFs in our calculations, we also provide the factorization $K$-factors $K$ and $K_{\text{PDF}}$ in these tables. The NLO+NLL QCD relative correction $\delta^{\text{NLO+NLL}}$ and the matrix-element-induced factorization $K$-factor $K_{\text{PME}}$ can be straightforwardly calculated by using $\delta^{\text{NLO}}$, $K$, and $K_{\text{PDF}}$,
\begin{equation}
\delta^{\text{NLO+NLL}}
=
\left( K - 1 \right) + K \delta^{\text{NLO}},
\qquad
\qquad
K_{\text{PME}} = \frac{K}{K_{\text{PDF}}}.
\end{equation}
As shown in Tables \ref{Xsection-0}, \ref{Xsection-50}, and \ref{Xsection-100}, the QCD correction can significantly enhance the LO production cross section, especially for light scalar-pseudoscalar pair. The NLO QCD relative correction exceeds $30\%$ at $m = 50~ \text{GeV}$, and decreases gradually to approximately $5.9\%$, $5.3\%$ and $4.8\%$ for $\Delta m = 0$, $50$, and $100~ \text{GeV}$, respectively, as $m$ increases to $800~ \text{GeV}$. The full NLL resummation correction (quantitatively described by $K-1$) slightly enhances the NLO QCD corrected cross section as $m < 500~ \text{GeV}$, but suppresses it by approximately $2\%$ in the high mass region. Compared to the full NLL resummation correction, the contribution from the threshold-resummation improved PDFs is more sensitive to $m$. The corresponding relative correction, i.e., $K_{\text{PDF}} - 1$, decreases monotonically as the increment of $m$, and reaches approximately $-5\%$ when $m = 800~ \text{GeV}$. Moreover, we find that the impact of the threshold-resummation improved PDFs becomes increasingly important with the increment of $m$ for heavy scalar-pseudoscalar pair. On the contrary, the NLL QCD relative correction from the partonic matrix element, $K_{\text{PME}} - 1$, increases monotonically with the increment of $m$. It is almost independent of the mass splitting between $\phi^0$ and $A^0$, varying from approximately $-1\%$ to $4\%$ as $m$ increases from $50$ to $800~ \text{GeV}$.
%-----------------------------------------------------Table 4-----------------------------------------------------
\begin{table}[!htbp]
\renewcommand \tabcolsep{7.0pt}
\renewcommand \arraystretch{1.1}
\centering
\begin{tabular}{|c|cccccc|}
\hline\hline
  $m~\text{[GeV]}$
& $\sigma^{\text{LO}}~\text{[fb]}$
& $\sigma^{\text{NLO}}~\text{[fb]}$
& $\sigma^{\text{NLO+NLL}}~\text{[fb]}$
& $\delta^{\text{NLO}}~ \text{[\%]}$
& $K$
& $K_{\text{PDF}}$
\\
\hline
  $50$
& $4923.7_{-9.4\%}^{+8.5\%}$ & $6717.8_{-0.4\%-3.7\%}^{+0.7\%+2.8\%}$ & $6731.0_{-0.6\%-3.7\%}^{+0.0\%+2.8\%}$
& $36.4$
& $1.002$ & $1.013$
\\
  $100$
& $218.7_{-3.2\%}^{+2.5\%}$  & $290.2_{-0.4\%-3.8\%}^{+0.9\%+2.9\%}$  & $290.8_{-0.4\%-3.8\%}^{+0.0\%+2.9\%}$
& $32.7$
& $1.002$ & $1.011$
\\
  $150$
& $49.41_{-0.5\%}^{+0.1\%}$  & $63.75_{-0.9\%-4.3\%}^{+1.4\%+3.3\%}$  & $63.94_{-0.4\%-4.3\%}^{+0.0\%+3.3\%}$
& $29.0$
& $1.003$  & $1.009$
\\
  $200$
& $16.98_{-1.4\%}^{+1.2\%}$  & $21.40_{-1.3\%-4.7\%}^{+1.6\%+3.6\%}$  & $21.49_{-0.6\%-4.7\%}^{+0.1\%+3.6\%}$
& $26.0$
& $1.004$ & $1.006$
\\
  $300$
& $3.504_{-3.3\%}^{+3.4\%}$  & $4.249_{-1.8\%-5.7\%}^{+1.9\%+4.3\%}$  & $4.267_{-1.2\%-5.7\%}^{+0.8\%+4.3\%}$
& $21.3$
& $1.004$ & $1.001$
\\
  $400$
& $1.051_{-4.5\%}^{+4.9\%}$  & $1.233_{-2.2\%-6.3\%}^{+2.2\%+5.0\%}$  & $1.237_{-1.7\%-6.3\%}^{+1.5\%+5.0\%}$
& $17.3$
& $1.003$ & $0.994$
\\
  $500$
& $0.3848_{-5.4\%}^{+6.0\%}$ & $0.4387_{-2.5\%-7.1\%}^{+2.4\%+5.8\%}$ & $0.4392_{-2.2\%-7.1\%}^{+2.1\%+5.8\%}$
& $14.0$
& $1.001$ & $0.985$
\\
  $600$
& $0.1597_{-6.2\%}^{+6.9\%}$ & $0.1773_{-2.8\%-7.9\%}^{+2.6\%+6.6\%}$ & $0.1771_{-2.9\%-7.9\%}^{+2.6\%+6.6\%}$
& $11.0$
& $0.999$ & $0.977$
\\
  $700$
& $0.07213_{-6.9\%}^{+7.8\%}$ & $0.07818_{-3.1\%-8.8\%}^{+2.8\%+7.5\%}$ & $0.07770_{-3.5\%-8.8\%}^{+3.2\%+7.5\%}$
& $8.39$
& $0.994$ & $0.966$
\\
  $800$
& $0.03465_{-7.4\%}^{+8.5\%}$ & $0.03670_{-3.3\%-9.7\%}^{+3.0\%+8.4\%}$ & $0.03618_{-4.4\%-9.7\%}^{+3.7\%+8.4\%}$
& $5.92$
& $0.986$ & $0.952$
\\
\hline\hline
\end{tabular}
\caption{
\label{Xsection-0}
LO, NLO, NLO+NLL QCD corrected integrated cross sections, NLO QCD relative corrections, and factorization $K$-factors ($K$ and $K_{\text{PDF}}$) for $pp \rightarrow q\bar{q} \rightarrow \phi^0A^0$ at $\sqrt{s} = 13~ \text{TeV}$ LHC within the 2HDM. The cross section central values are folded with the theoretical relative errors estimated from scale variation (first quote) and PDFs (second quote). The mass splitting between $\phi^0$ and $A^0$ is fixed to zero ($\Delta m = 0$).}
\end{table}
%-----------------------------------------------------------------------------------------------------------------
%-----------------------------------------------------Table 5-----------------------------------------------------
\begin{table}[!htbp]
\renewcommand \tabcolsep{7.0pt}
\renewcommand \arraystretch{1.1}
\centering
\begin{tabular}{|c|cccccc|}
\hline\hline
  $m~\text{[GeV]}$
& $\sigma^{\text{LO}}~\text{[fb]}$
& $\sigma^{\text{NLO}}~\text{[fb]}$
& $\sigma^{\text{NLO+NLL}}~\text{[fb]}$
& $\delta^{\text{NLO}}~ \text{[\%]}$
& $K$
& $K_{\text{PDF}}$
\\
\hline
  $50$
& $611.1_{-5.3\%}^{+4.5\%}$ & $824.5_{-0.0\%-3.7\%}^{+0.5\%+2.8\%}$ & $825.8_{-0.4\%-3.7\%}^{+0.0\%+2.8\%}$
& $34.9$
& $1.002$ & $1.012$
\\
  $100$
& $93.88_{-1.6\%}^{+1.1\%}$ & $122.7_{-0.7\%-4.0\%}^{+1.2\%+3.1\%}$ & $123.0_{-0.3\%-4.0\%}^{+0.0\%+3.1\%}$
& $30.7$
& $1.002$ & $1.010$
\\
  $150$
& $27.61_{-0.7\%}^{+0.4\%}$ & $35.19_{-1.1\%-4.5\%}^{+1.5\%+3.5\%}$ & $35.31_{-0.5\%-4.5\%}^{+0.0\%+3.5\%}$
& $27.5$
& $1.003$ & $1.007$
\\
  $200$
& $10.77_{-2.0\%}^{+1.9\%}$ & $13.43_{-1.4\%-4.9\%}^{+1.7\%+3.8\%}$ & $13.48_{-0.8\%-4.9\%}^{+0.3\%+3.8\%}$
& $24.7$
& $1.004$ & $1.005$
\\
  $300$
& $2.517_{-3.6\%}^{+3.8\%}$ & $3.026_{-1.9\%-5.8\%}^{+2.0\%+4.5\%}$ & $3.038_{-1.3\%-5.8\%}^{+1.0\%+4.5\%}$
& $20.2$
& $1.004$ & $0.999$
\\
  $400$
& $0.8032_{-4.8\%}^{+5.2\%}$ & $0.9353_{-2.3\%-6.5\%}^{+2.2\%+5.2\%}$ & $0.9388_{-1.9\%-6.5\%}^{+1.6\%+5.2\%}$
& $16.4$
& $1.004$ & $0.992$
\\
  $500$
& $0.3053_{-5.6\%}^{+6.2\%}$ & $0.3457_{-2.6\%-7.3\%}^{+2.4\%+6.0\%}$ & $0.3458_{-2.5\%-7.3\%}^{+2.2\%+6.0\%}$
& $13.2$
& $1.000$ & $0.983$
\\
  $600$
& $0.1299_{-6.4\%}^{+7.2\%}$ & $0.1433_{-2.8\%-8.1\%}^{+2.7\%+6.8\%}$ & $0.1430_{-3.1\%-8.1\%}^{+2.8\%+6.8\%}$
& $10.3$
& $0.998$ & $0.974$
\\
  $700$
& $0.05969_{-7.0\%}^{+7.9\%}$ & $0.06432_{-3.1\%-9.0\%}^{+2.9\%+7.7\%}$ & $0.06361_{-3.3\%-9.0\%}^{+3.7\%+7.7\%}$
& $7.76$
& $0.989$ & $0.960$
\\
  $800$
& $0.02904_{-7.6\%}^{+8.6\%}$ & $0.03059_{-3.4\%-9.9\%}^{+3.1\%+8.7\%}$ & $0.03008_{-4.4\%-9.9\%}^{+3.9\%+8.7\%}$
& $5.34$
& $0.983$ & $0.949$
\\
\hline\hline
\end{tabular}
\caption{
\label{Xsection-50}
Same as Table \ref{Xsection-0} but for $\Delta m = 50~ \text{GeV}$.}
\end{table}
%-----------------------------------------------------------------------------------------------------------------
%-----------------------------------------------------Table 6-----------------------------------------------------
\begin{table}[!htbp]
\renewcommand \tabcolsep{6.0pt}
\renewcommand \arraystretch{1.1}
\centering
\begin{tabular}{|c|cccccc|}
\hline\hline
  $m~\text{[GeV]}$
& $\sigma^{\text{LO}}~\text{[fb]}$
& $\sigma^{\text{NLO}}~\text{[fb]}$
& $\sigma^{\text{NLO+NLL}}~\text{[fb]}$
& $\delta^{\text{NLO}}~ \text{[\%]}$
& $K$
& $K_{\text{PDF}}$
\\
\hline
  $50$
& $185.0_{-3.0\%}^{+2.4\%}$ & $245.3_{-0.4\%-3.9\%}^{+1.0\%+2.9\%}$ & $245.7_{-0.4\%-3.9\%}^{+0.0\%+2.9\%}$
& $32.6$
& $1.002$ & $1.011$
\\
  $100$
& $45.97_{-0.4\%}^{+0.1\%}$ & $59.26_{-1.0\%-4.3\%}^{+1.4\%+3.3\%}$ & $59.43_{-0.4\%-4.3\%}^{+0.0\%+3.3\%}$
& $28.9$
& $1.003$ & $1.009$
\\
  $150$
& $16.30_{-1.4\%}^{+1.2\%}$ & $20.53_{-1.4\%-4.8\%}^{+1.5\%+3.6\%}$ & $20.60_{-0.7\%-4.8\%}^{+0.1\%+3.6\%}$
& $26.0$
& $1.003$ & $1.006$
\\
  $200$
& $7.033_{-2.5\%}^{+2.4\%}$ & $8.683_{-1.6\%-5.2\%}^{+1.8\%+4.0\%}$ & $8.717_{-0.9\%-5.2\%}^{+0.5\%+4.0\%}$
& $23.5$
& $1.004$ & $1.004$
\\
  $300$
& $1.832_{-3.9\%}^{+4.2\%}$ & $2.183_{-2.0\%-5.9\%}^{+2.1\%+4.7\%}$ & $2.192_{-1.5\%-5.9\%}^{+1.1\%+4.7\%}$
& $19.2$
& $1.004$ & $0.998$
\\
  $400$
& $0.6183_{-5.0\%}^{+5.5\%}$ & $0.7146_{-2.3\%-6.7\%}^{+2.3\%+5.4\%}$ & $0.7168_{-2.1\%-6.7\%}^{+1.8\%+5.4\%}$
& $15.6$
& $1.003$ & $0.990$
\\
  $500$
& $0.2433_{-5.8\%}^{+6.5\%}$ & $0.2736_{-2.6\%-7.5\%}^{+2.5\%+6.2\%}$ & $0.2738_{-2.7\%-7.5\%}^{+2.3\%+6.2\%}$
& $12.5$
& $1.001$ & $0.982$
\\
  $600$
& $0.1059_{-6.5\%}^{+7.4\%}$ & $0.1161_{-2.9\%-8.4\%}^{+2.7\%+7.1\%}$ & $0.1157_{-3.2\%-8.4\%}^{+2.9\%+7.1\%}$
& $9.63$
& $0.997$ & $0.972$
\\
  $700$
& $0.04949_{-7.2\%}^{+8.1\%}$ & $0.05301_{-3.2\%-9.2\%}^{+2.9\%+7.9\%}$ & $0.05250_{-3.8\%-9.2\%}^{+3.4\%+7.9\%}$
& $7.11$
& $0.990$ & $0.959$
\\
  $800$
& $0.02438_{-7.7\%}^{+8.8\%}$
& $~0.02554_{-3.4\%-10.1\%}^{+3.1\%+9.0\%}$
& $~0.02505_{-4.4\%-10.1\%}^{+3.9\%+9.0\%}$
& $4.76$
& $0.981$ & $0.945$
\\
\hline\hline
\end{tabular}
\caption{
\label{Xsection-100}
Same as Table \ref{Xsection-0} but for $\Delta m = 100~ \text{GeV}$.}
\end{table}
%-----------------------------------------------------------------------------------------------------------------

\par
The QCD production of $\phi^0A^0$ via gluon-gluon fusion depends not only on $m_{\phi^0}$ and $m_{A^0}$, but also on $m_{12}^2$ and $\tan\beta$, since the Yukawa couplings and triple Higgs self-couplings are involved in this production channel. We calculate the lowest-order production cross section for this loop-induced channel at the two benchmark points listed in Table \ref{benchmarks}, which can satisfy both theoretical and experimental constraints. At both benchmark points, $H^0$ is the BSM $\mathcal{CP}$-even Higgs boson, i.e., $H^0 = \phi^0$, and $\sin(\beta - \alpha) = 1$ at the alignment limit. The other two Higgs parameters of 2HDM, $m_{h^0}$ and $m_{H^{\pm}}$, are not given in Table \ref{benchmarks}, because the scalar-pseudoscalar pair production at QCD NLO+NLL accuracy is completely independent from the SM-like and charged Higgs bosons. The integrated cross sections for $pp \rightarrow gg \rightarrow \phi^0A^0$ at $\sqrt{s} = 13~ \text{TeV}$ LHC in the type-I 2HDM at $\text{BP1}$ and $\text{BP2}$, listed in Table \ref{Xsection-gg}, are approximately $1$ and $0.06~ \text{fb}$, respectively. We can see that $\sigma_{gg}/\sigma^{\text{NLO+NLL}}$, i.e., the ratio of the contribution from gluon-gluon fusion channel to the NLO+NLL QCD corrected cross section of quark-antiquark annihilation channel, is approximately $2.9\%$ at $\text{BP1}$ and can reach $8.0\%$ at $\text{BP2}$. It can be concluded that the scalar-pseudoscalar pair production at the LHC in the type-I 2HDM is predominated by the quark-initiated Drell-Yan production channel, and the gluon-gluon fusion contribution is non-negligible and should be taken into consideration in precision predictions.
%-----------------------------------------------------Table 7-----------------------------------------------------
\begin{table}[!htbp]
\renewcommand \tabcolsep{8.0pt}
\centering
\begin{tabular}{|c|cccc|}
\hline\hline
Benchmark point & $m_{H^0}~\text{[GeV]}$ & $m_{A^0}~\text{[GeV]}$ & $m_{12}^2$ & $\tan\beta$  \\
\hline
BP1 & 150 & 200 &  2000 & 10 \\
BP2 & 400 & 500 & 50000 &  2 \\
\hline\hline
\end{tabular}
\caption{
\label{benchmarks} Benchmark points BP1 and BP2.}
\end{table}
%-----------------------------------------------------------------------------------------------------------------
%-----------------------------------------------------Table 8-----------------------------------------------------
\begin{table}[!htbp]
\renewcommand \tabcolsep{10.0pt}
\renewcommand \arraystretch{1.1}
\centering
\begin{tabular}{|c|cc|}
\hline\hline
  Benchmark point
& BP1 & BP2 \\
\hline
  $\sigma_{gg}~\text{[fb]}$
& $1.020_{-19.8\%-3.4\%}^{+26.4\%+3.7\%}$ & $0.05742^{+31.8\%+8.0\%}_{-22.7\%-6.4\%}$
\\
\hline\hline
\end{tabular}
\caption{
\label{Xsection-gg}
Lowest-order integrated cross sections for $pp \rightarrow gg \rightarrow \phi^0A^0$ at $\sqrt{s} = 13~ \text{TeV}$ LHC in type-I 2HDM at the benchmark points BP1 and BP2 .}
\end{table}
%-----------------------------------------------------------------------------------------------------------------

\subsection{Transverse momentum distribution}
\par
\begin{figure}[htb]
\centering
\includegraphics[width=0.48\textwidth]{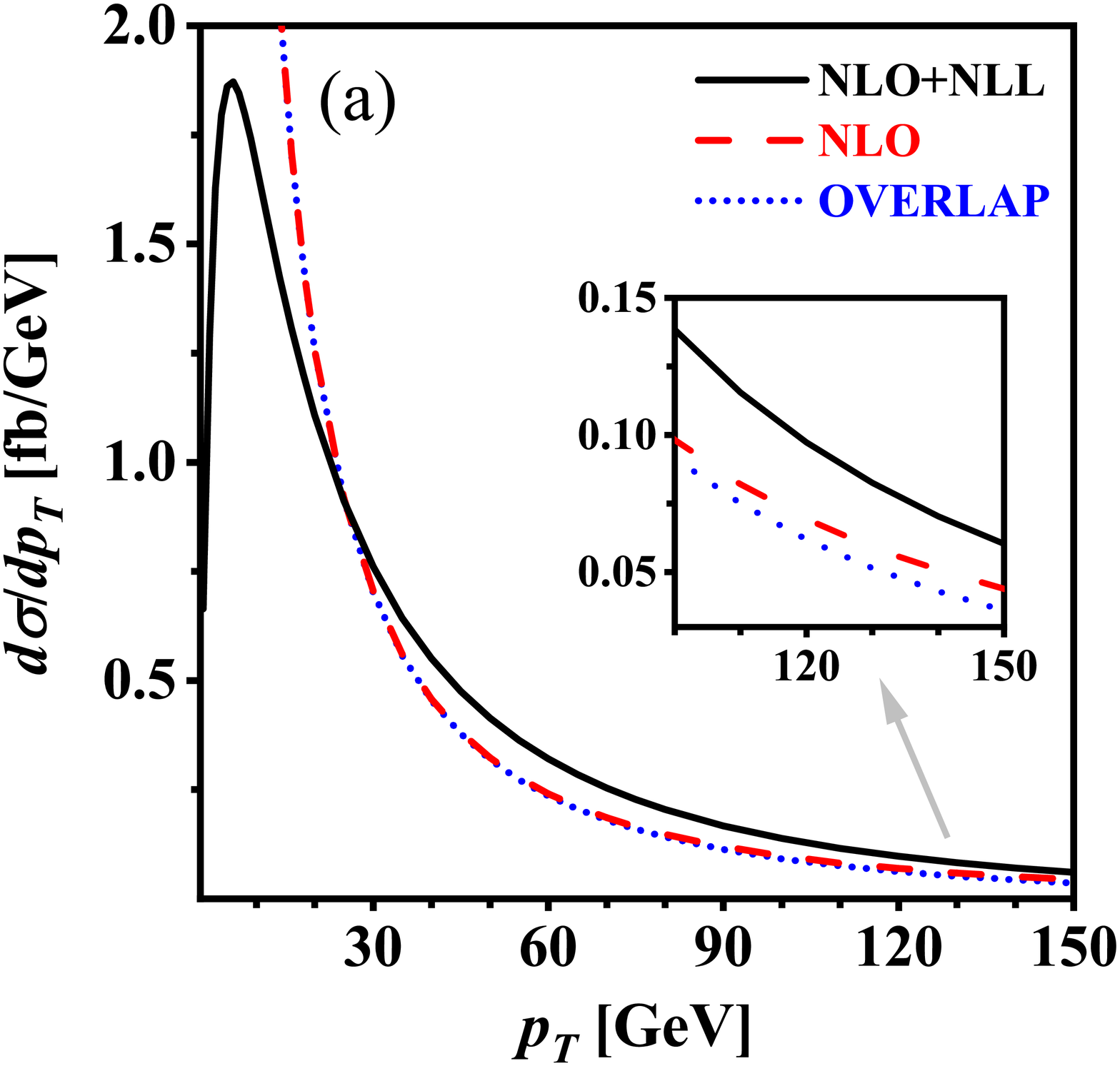}
\includegraphics[width=0.48\textwidth]{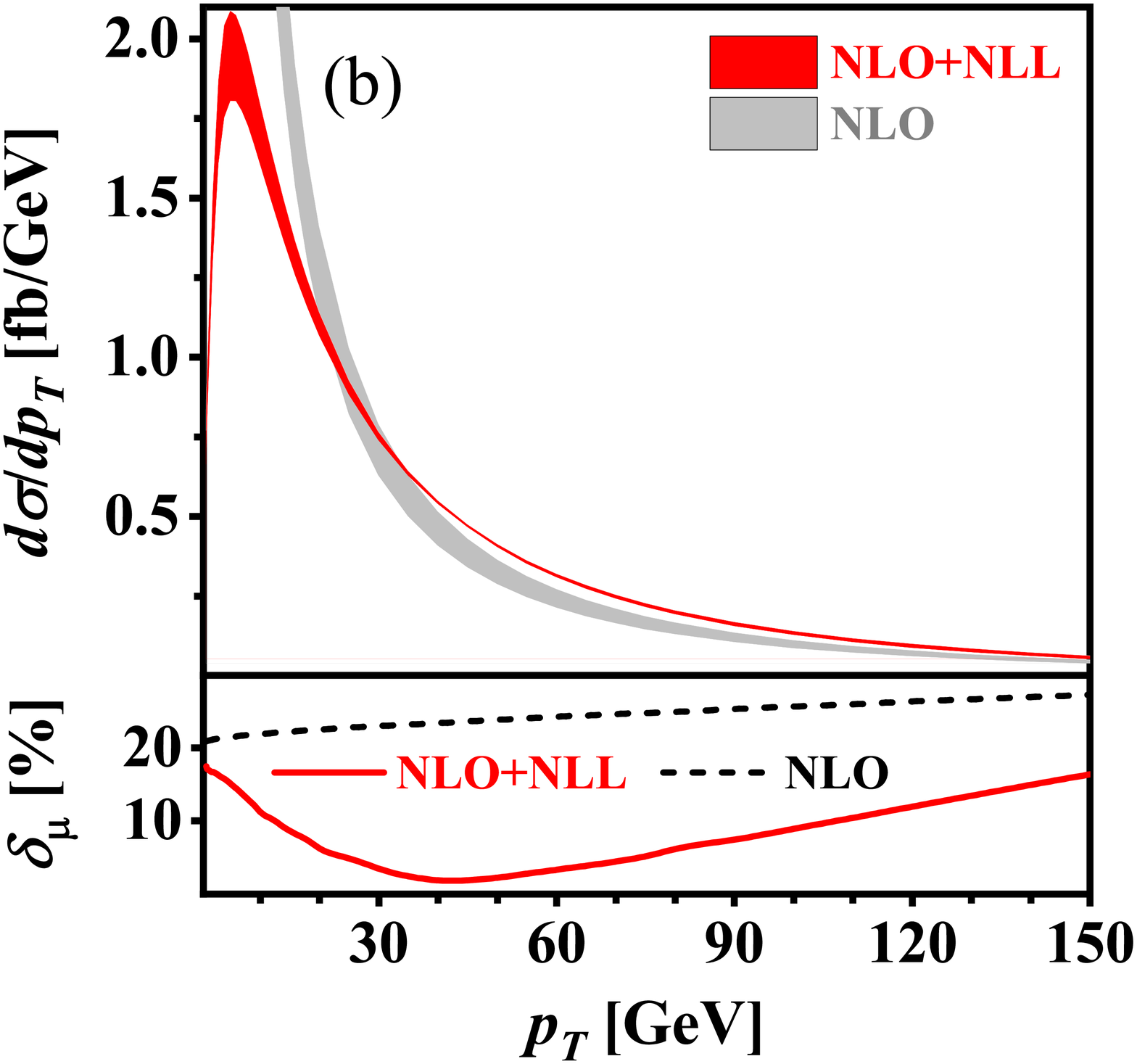}
\centering \caption{
\label{pT-BP1}
(a) Transverse momentum distribution of final-state $\phi^0A^0$ and (b) its scale uncertainty for $pp \rightarrow q\bar{q} \rightarrow \phi^0A^0$ at $\sqrt{s} = 13~ \text{TeV}$ LHC within the 2HDM at the benchmark point $\text{BP1}$.}
\end{figure}
\begin{figure}[htb]
\centering
\includegraphics[width=0.48\textwidth]{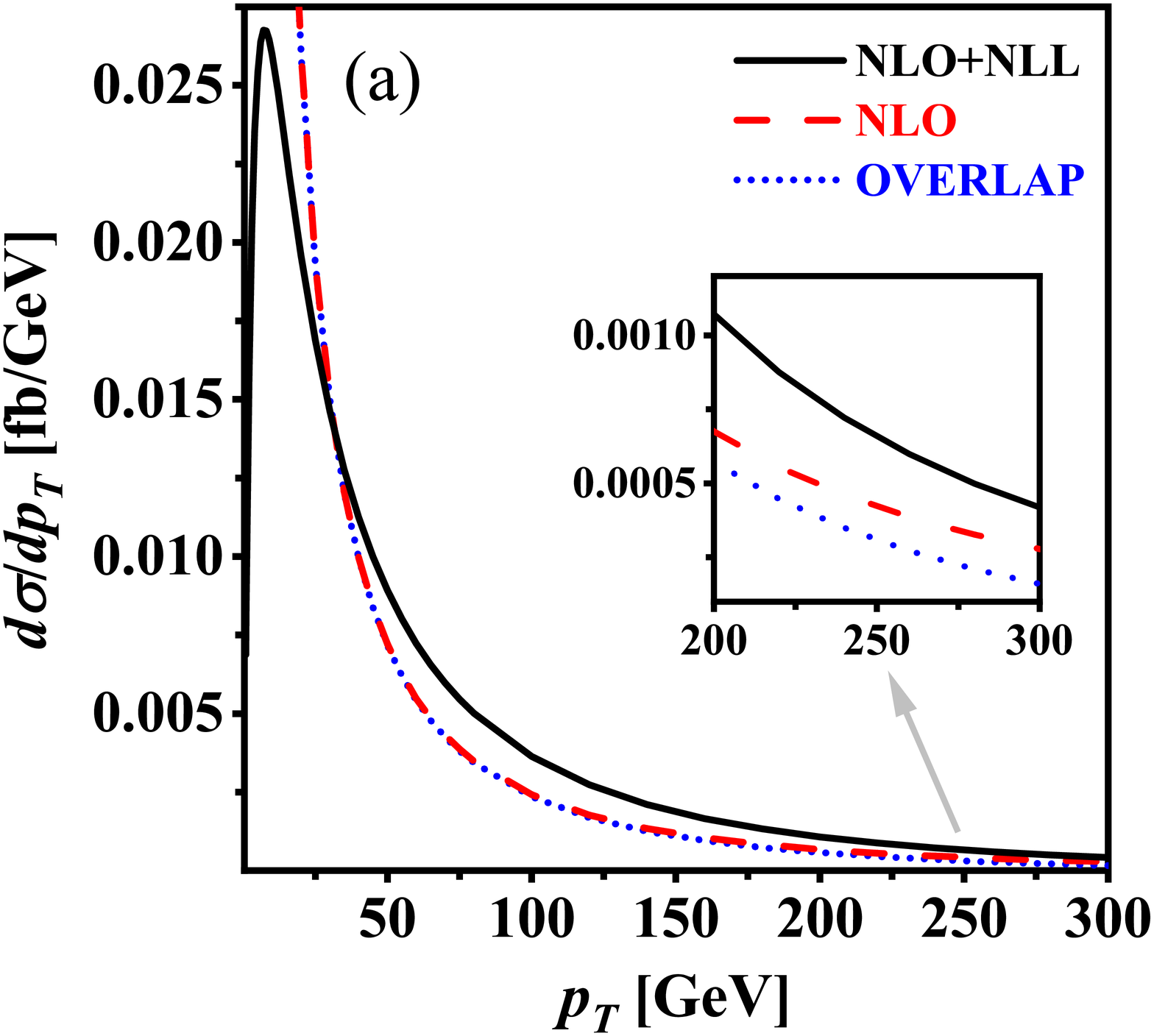}
\includegraphics[width=0.48\textwidth]{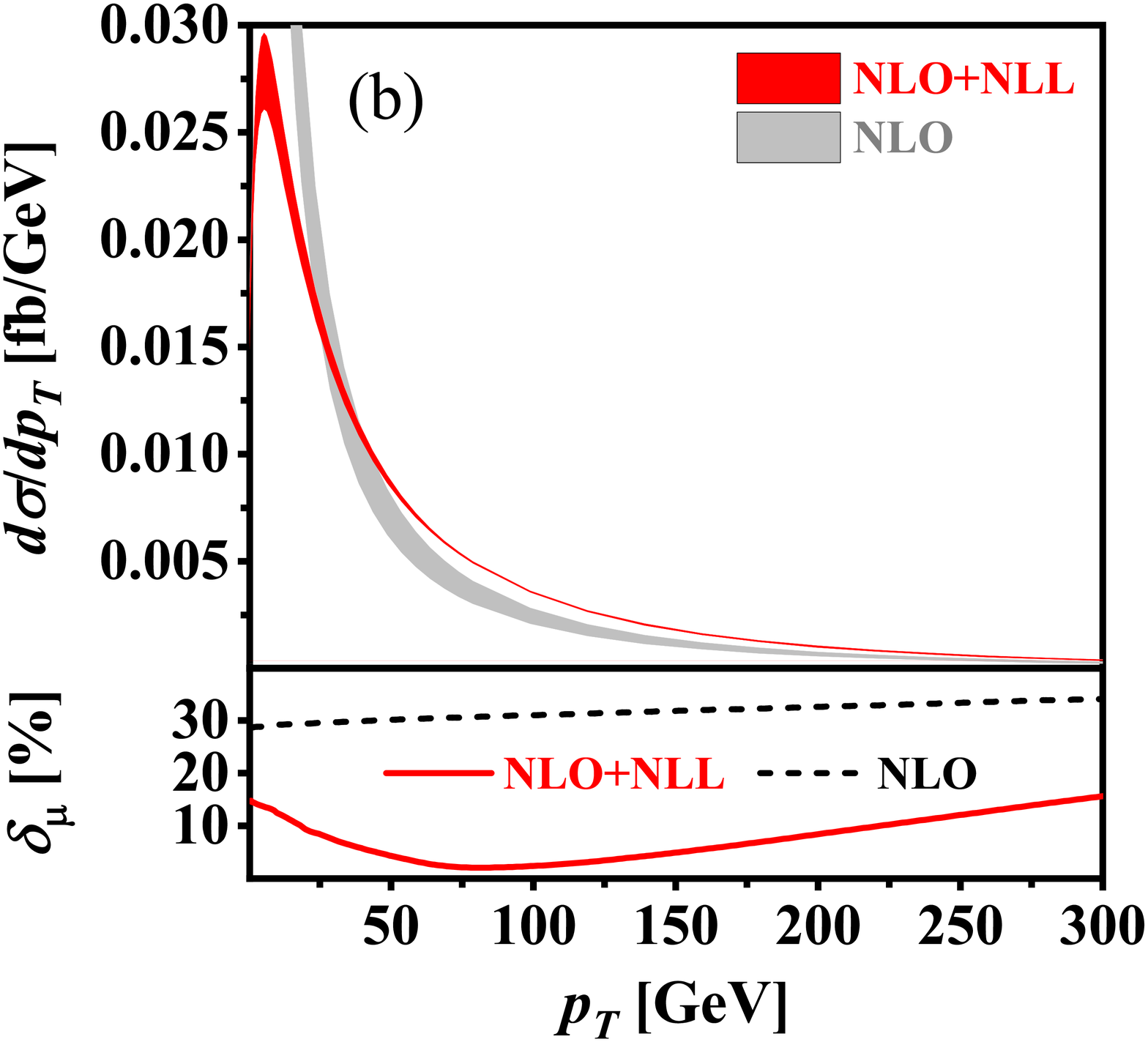}
\centering
\caption{
\label{pT-BP2}
Same as Fig.\ref{pT-BP1} but at $\text{BP2}$.}
\end{figure}

Next, we address the transverse momentum distribution of the scalar-pseudoscalar pair produced at the LHC. Since the one-loop-induced gluon-gluon fusion channel does not contribute to the $p_T$ distribution due to the momentum conservation, we consider only the quark-initiated Drell-Yan production channel. The NLO, NLO+NLL QCD corrected transverse momentum distributions of $\phi^0A^0$ as well as the overlap between the NLO QCD corrected and NLL QCD resummed $p_T$ distributions (labeled by ``NLO", ``NLO+NLL", and ``OVERLAP") for the Drell-Yan production of $\phi^0A^0$ at $\sqrt{s} = 13~ \text{TeV}$ LHC in the 2HDM at the benchmark points $\text{BP1}$ and $\text{BP2}$ are shown in Figs.\ref{pT-BP1}(a) and \ref{pT-BP2}(a), respectively. The central scale is $\mu_0 = m_{\phi^0} + m_{A^0}$. As expected, the NLO QCD corrected $p_T$ distribution and the overlap $p_T$ distribution are in good agreement with each other in the small-$p_T$ region \cite{Bozzi:2005wk} and become divergent as $p_T \rightarrow 0$, but the discrepancy between them becomes increasingly evident with the increment of $p_T$. The relative discrepancy between the NLO QCD corrected and the overlap $p_T$ distributions, defined as
\begin{equation}
\eta
=
\left(
\frac{d \sigma^{\text{NLO}}}{d p_T} - \frac{d \sigma^{\text{OVERLAP}}}{d p_T}
\right)
\left/
\frac{d \sigma^{\text{NLO}}}{d p_T}
\right.,
\end{equation}
can reach about $18.2\%$ and $42.6\%$ when $p_T = 150$ and $300~ \text{GeV}$ at $\text{BP1}$ and $\text{BP2}$, respectively. Compared to the NLO QCD corrected $p_T$ distribution, the NLO+NLL QCD corrected $p_T$ distribution is finite and more reliable in the whole final-state phase space. It increases sharply in the small-$p_T$ region, reaches its maximum of about $1.9~ \text{fb/GeV}$ in the vicinity of $p_T \sim 5.5~ \text{GeV}$, and then decreases approximately logarithmically as the increment of $p_T$ at the benchmark point $\text{BP1}$. As for the benchmark point $\text{BP2}$, the NLO+NLL QCD corrected $p_T$ distribution peaks at $p_T \sim 7.5~ \text{GeV}$ and its maximum is about $0.027~ \text{fb/GeV}$.

\par
The scale uncertainty of a differential distribution with respect to some kinematic variable $x$ can be estimated by
\begin{equation}
\delta_{\mu}(x)
=
\max
\left\{
\frac{d \sigma}{d x}(\mu_1) - \frac{d \sigma}{d x}(\mu_2)
\right\}
\left/
\frac{d \sigma}{d x}(\mu_0)
\right.,
\qquad
\mu_1,\, \mu_2 \in \left[ \mu_0/2,~ 2\mu_0 \right]
\end{equation}
In Figs.\ref{pT-BP1}(b) and \ref{pT-BP2}(b), we plot the scale uncertainties of the NLO and NLO+NLL QCD corrected $p_T$ distributions, denoted by $\delta_{\mu}^{\text{NLO}}$ and $\delta_{\mu}^{\text{NLO+NLL}}$, at $\text{BP1}$ and $\text{BP2}$, respectively. As shown in the lower panels of Figs.\ref{pT-BP1}(b) and \ref{pT-BP2}(b), the scale uncertainty of the NLO QCD corrected $p_T$ distribution increases gradually, while the scale uncertainty of the NLO+NLL QCD corrected $p_T$ distribution first decreases consistently before reaching its minimum and then increases monotonically, as the increment of $p_T$. Some representative values of $\delta_{\mu}^{\text{NLO}}$ and $\delta_{\mu}^{\text{NLO+NLL}}$ are given in Table \ref{delta-mu-pt}. This table, as well as Figs.\ref{pT-BP1}(b) and \ref{pT-BP2}(b), clearly shows that $\delta_{\mu}^{\text{NLO+NLL}}$ is much less than $\delta_{\mu}^{\text{NLO}}$, especially in the intermediate-$p_T$ region. Thus, we conclude that the resummation of higher-order large logarithmic contributions can significantly improve the fixed-order prediction for the $p_T$ distribution; the NLO+NLL QCD corrected $p_T$ distribution is much more reliable in the whole $p_T$ region compared to the NLO QCD corrected $p_T$ distribution.
%-----------------------------------------------------Figure 3-----------------------------------------------------
%------------------------------------------------------------------------------------------------------------------
%-----------------------------------------------------Figure 4-----------------------------------------------------
%------------------------------------------------------------------------------------------------------------------
%-----------------------------------------------------Table 9-----------------------------------------------------
\begin{table}[!htbp]
\renewcommand \tabcolsep{10.0pt}
\centering
\begin{tabular}{|c|c|c|c|c|}
\hline\hline
\multirow{2}*{Benchmark point}
& \multicolumn{2}{c|}{BP1} & \multicolumn{2}{c|}{BP2} \\
\cline{2-5}
& $p_T = 1~ \text{GeV}$ & $p_T = 150~ \text{GeV}$ & $p_T = 2~ \text{GeV}$ & $p_T = 300~ \text{GeV}$ \\
\hline
$\delta_{\mu}^{\text{NLO}}$                    & $20.9\%$ & $27.2\%$ & $28.7\%$ & $34.0\%$ \\
\hline
\multirow{2}*{$\delta_{\mu}^{\text{NLO+NLL}}$} & $17.2\%$ & $16.3\%$ & $16.9\%$ & $15.7\%$ \\
\cline{2-5}
& \multicolumn{2}{c|}{$\text{min.} \simeq 1.6\%_{\quad \left( @\, p_T \,\sim\, 45\, \text{GeV} \right)}$}
& \multicolumn{2}{c|}{$\text{min.} \simeq 2.0\%_{\quad \left( @\, p_T \,\sim\, 80\, \text{GeV} \right)}$} \\
\hline\hline
\end{tabular}
\caption{
\label{delta-mu-pt}
Scale uncertainties of NLO and NLO+NLL QCD corrected $p_T$ distributions for $pp \rightarrow q\bar{q} \rightarrow \phi^0A^0$ at $\sqrt{s} = 13~ \text{TeV}$ LHC within the 2HDM at $\text{BP1}$ and $\text{BP2}$ for some typical values of $p_T$.}
\end{table}
%-----------------------------------------------------------------------------------------------------------------

\subsection{Invariant mass distribution}
\par
\begin{figure}[htb]
\centering
\includegraphics[width=0.48\textwidth]{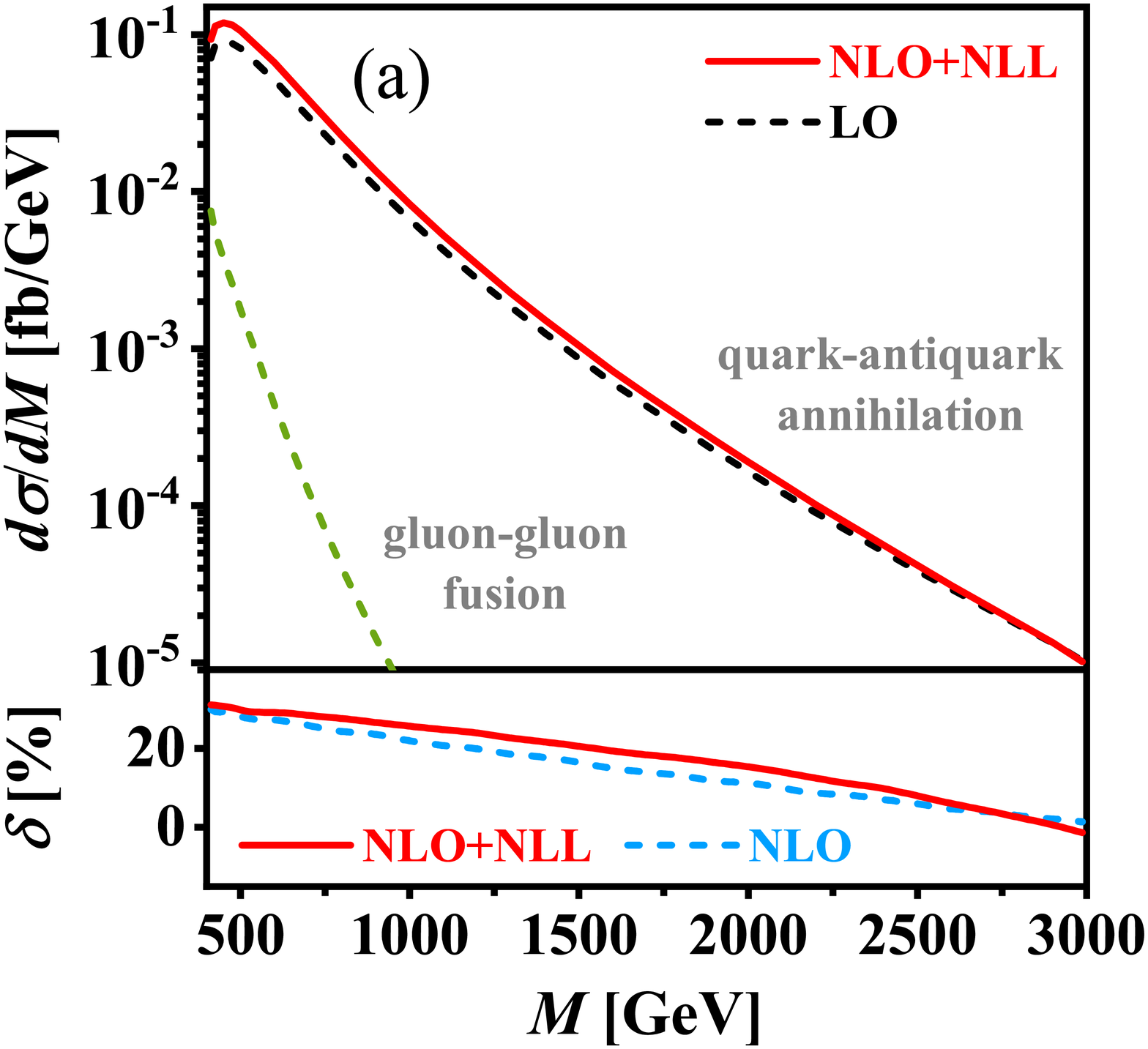}
\includegraphics[width=0.48\textwidth]{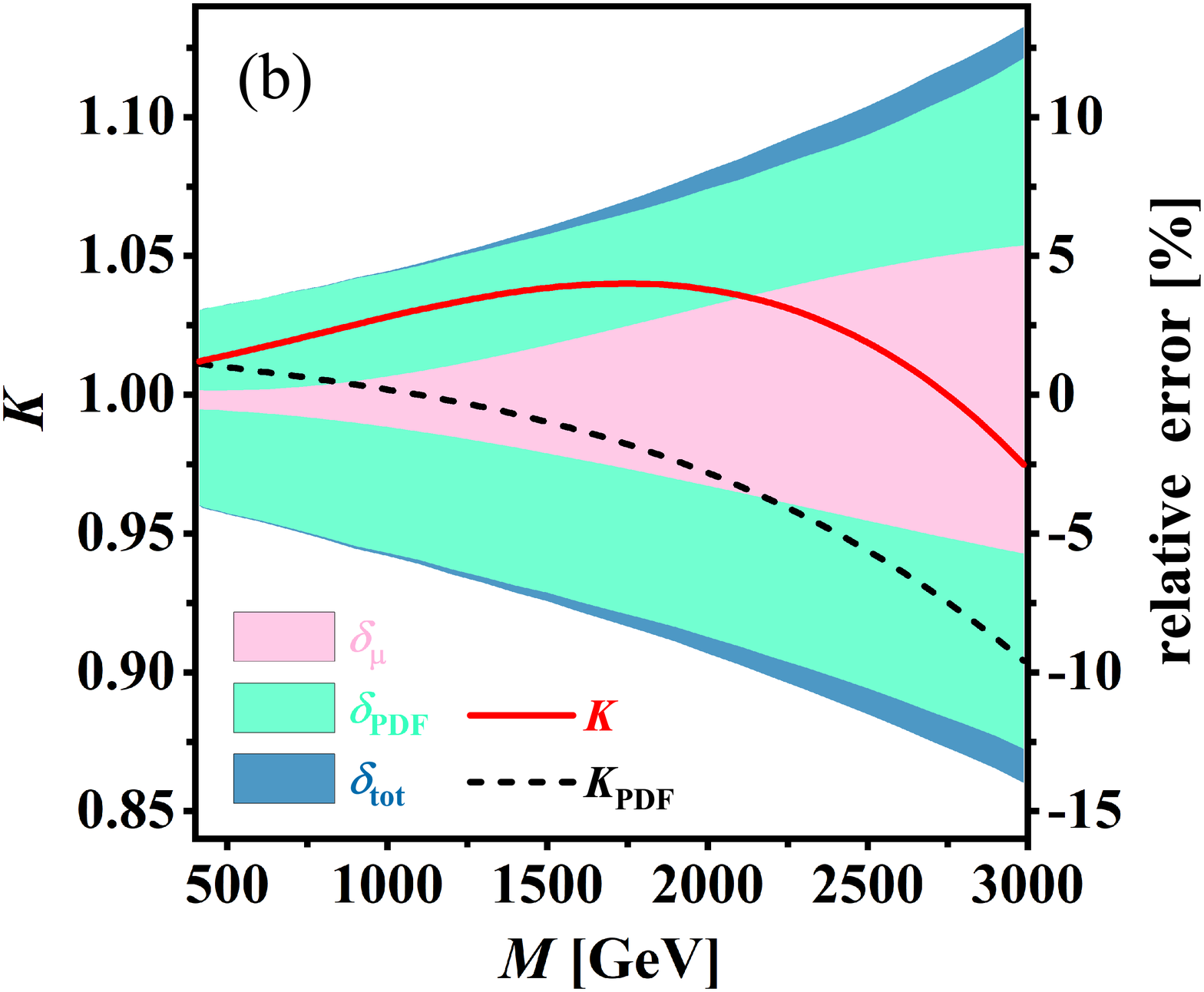}
\caption{
\label{M-BP1}
(a) Invariant mass distribution of final-state $\phi^0A^0$ and (b) factorization $K$-factors ($K$ and $K_{\text{PDF}}$) as well as theoretical relative errors ($\delta_{\mu}$, $\delta_{\text{PDF}}$ and $\delta_{\text{tot}}$) for $\phi^0A^0$ associated production at $\sqrt{s} = 13~ \text{TeV}$ LHC in type-I 2HDM at the benchmark point $\text{BP1}$.}
\end{figure}
\begin{figure}[htb]
\centering
\includegraphics[width=0.48\textwidth]{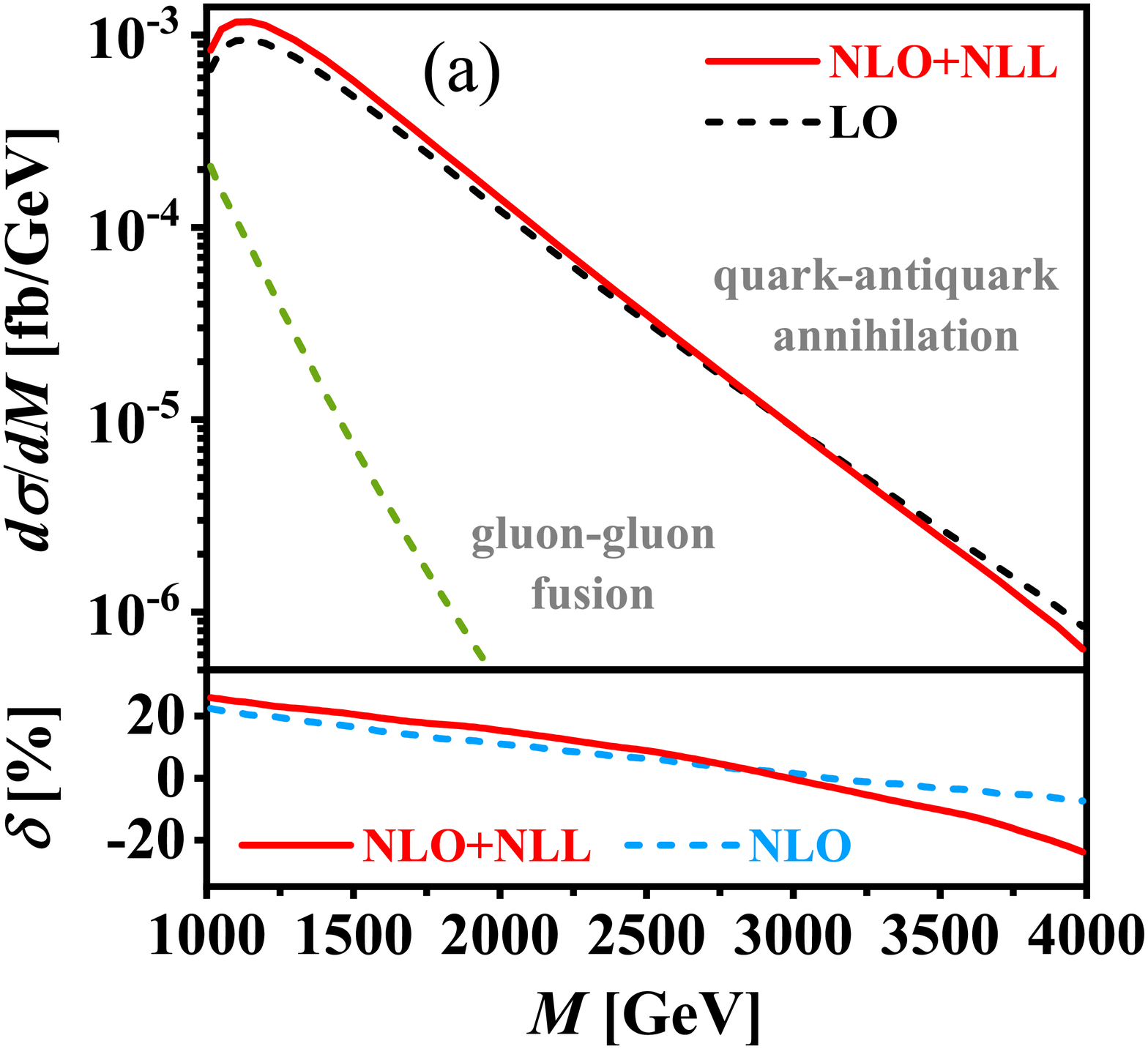}
\includegraphics[width=0.48\textwidth]{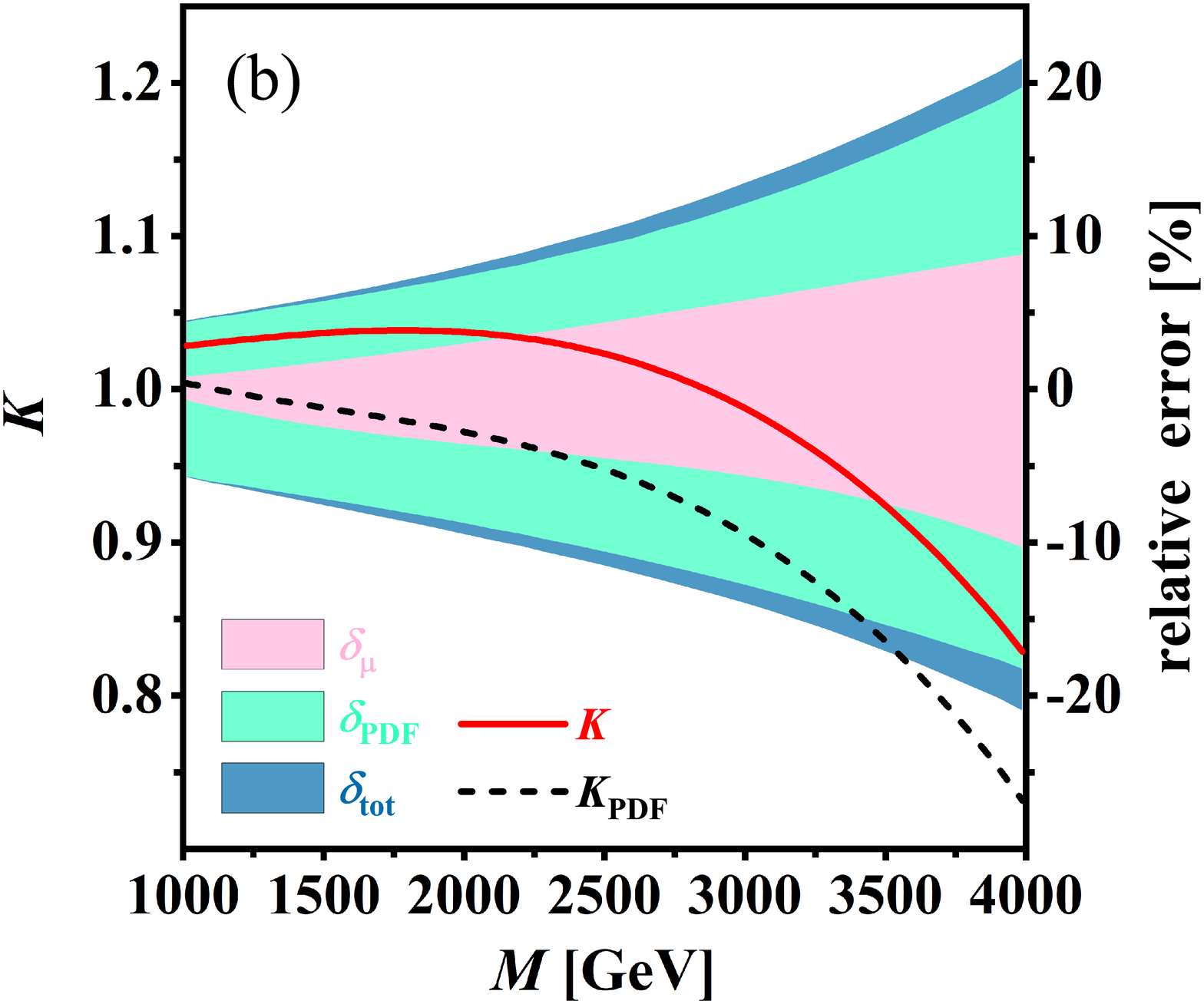}
\caption{
\label{M-BP2}
Same as Fig.\ref{M-BP1} but at $\text{BP2}$.}
\end{figure}

In this subsection, we discuss the threshold resummation effect on the invariant mass distribution of the scalar-pseudoscalar pair produced at the $13~ \text{TeV}$ LHC in the type-I 2HDM. The central scale is set to the invariant mass of the final-state scalar-pseudoscalar pair, i.e., $\mu_0 = M$. In the upper panels of Figs.\ref{M-BP1}(a) and \ref{M-BP2}(a), we depict the invariant mass distributions of the $\phi^0A^0$ system for both quark-initiated electroweak Drell-Yan production and gluon-initiated QCD production of $\phi^0A^0$ at $\text{BP1}$ and $\text{BP2}$, respectively. The corresponding NLO and NLO+NLL QCD relative corrections to the Drell-Yan production channel are provided in the lower panels. The $\phi^0A^0$ invariant mass distribution of the Drell-Yan channel increases rapidly near the production threshold, and then decreases consistently after reaching its maximum, as the increment of $M$. It peaks at $M \sim 450~ \text{GeV}$ for $\text{BP1}$ and $M \sim 1150~ \text{GeV}$ for $\text{BP2}$, respectively, at both LO and NLO+NLL accuracies. Compared to the Drell-Yan channel, the $\phi^0A^0$ invariant mass distribution of the gluon-gluon fusion channel is much smaller, and decreases more quickly as the increment of $M$. The ratio of the differential cross sections of the two channels, $d\sigma_{gg}/d\sigma^{\text{NLO+NLL}}$, is about $8.1\%$ at $M = 400~ \text{GeV}$ for $\text{BP1}$ and $24.9\%$ at $M = 1000~ \text{GeV}$ for $\text{BP2}$, respectively, and approaches zero rapidly as the increasing of $M$. It implies that the contribution from the gluon-gluon fusion channel is indispensable near the production threshold, but negligible in the high invariant mass region. The NLO and NLO+NLL QCD relative corrections ($\delta^{\text{NLO}}$ and $\delta^{\text{NLO+NLL}}$) to the Drell-Yan channel decrease gradually with the increment of $M$. They decrease from $29.8\%$ to $1.4\%$ and from $31.0\%$ to $-1.3\%$, respectively, as $M$ increases from $400~ \text{GeV}$ to $3~ \text{TeV}$ at $\text{BP1}$, and vary correspondingly in the range of $-7.3\% \sim 22.4\%$ and $-23.9\% \sim 25.9\%$ as $M \in [1,\, 4]~ \text{TeV}$ at $\text{BP2}$.

\par
To further demonstrate the full NLL resummation effect and the impact of the threshold-resummation improved PDFs on the $\phi^0A^0$ invariant mass distribution of the Drell-Yan channel, we plot the factorization $K$-factors $K$ and $K_{\text{PDF}}$ as functions of $M$ in Figs.\ref{M-BP1}(b) and \ref{M-BP2}(b) for $\text{BP1}$ and $\text{BP2}$, respectively. The theoretical errors from scale variation and PDFs as well as their combination, i.e., $\delta_{\mu}$, $\delta_{\text{PDF}}$ and $\delta_{\text{tot}}$, are also displayed in these two figures. At the benchmark point $\text{BP1}$, $K$ increases slowly from $1.01$ to $1.04$ as the increment of $M$ from $400~ \text{GeV}$ to $1.7~ \text{TeV}$, and then gradually decreases to $0.97$ as $M$ increases to $3~ \text{TeV}$. The full NLL resummation correction enhances the NLO QCD corrected invariant mass distribution of $\phi^0A^0$ in the region of $M < 2.7~ \text{TeV}$, but it would reduce the invariant mass distribution at sufficiently high invariant mass. However, $K_{\text{PDF}}$, which quantitatively reflects the impact of the threshold-resummation improved PDFs, decreases consistently from $1.01$ to $0.90$ as $M$ varies from $400~ \text{GeV}$ to $3~ \text{TeV}$. As for $K_{\text{PME}}$, which describes the NLL resummation effect from the partonic matrix element and is calculated by $K/K_{\text{PDF}}$, it shows the opposite tendency compared to $K_{\text{PDF}}$: it increases monotonically from $1.00$ to $1.08$ as the increment of $M$. At the benchmark point $\text{BP2}$, $K$ is fairly stable in the range of $1~ \text{TeV} < M < 2~ \text{TeV}$; it reaches the maximum of around $1.04$ at $M \sim 1.8~ \text{TeV}$ and subsequently decreases to $0.83$ as $M$ increases to $4~ \text{TeV}$. Simultaneously, a global suppression induced by the threshold-resummation improved PDFs can be clearly observed in the invariant mass distribution. Such suppression effect is very small and could be neglected at relatively low invariant mass, but becomes more and more apparent as the increasing of $M$. At $M = 4~ \text{TeV}$, $K_{\text{PDF}} = 0.73$; the contribution from the threshold-resummation improved PDFs is more considerable at high invariant mass compared to the NLO QCD correction. On the contrary, $K_{\text{PME}}$ increases from $1.02$ to $1.13$ as $M$ increases from $1$ to $4~ \text{TeV}$. In the high invariant mass region, the contribution from the threshold-resummation improved PDFs is the dominant correction compared to the NLO QCD correction and the NLL resummation correction from the partonic matrix element. For example, at $M = 4~ \text{TeV}$, $K_{\text{PDF}} - 1 = -27\%$, $\delta^{\text{NLO}} = -7.3\%$ and $K_{\text{PME}} - 1 = 13\%$, respectively.
%-----------------------------------------------------Figure 5-----------------------------------------------------
%------------------------------------------------------------------------------------------------------------------
%-----------------------------------------------------Figure 6-----------------------------------------------------
%------------------------------------------------------------------------------------------------------------------

\section{Summary}
\label{section-5}
\par
Searching for BSM Higgs bosons is an important task at the LHC and future high-energy colliders. In this study, we comprehensively analyze the scalar-pseudoscalar pair production at the $13~ \text{TeV}$ LHC at the alignment limit in the type-I 2HDM. The Collins-Soper-Sterman resummation approach and the factorization method are employed to resum the NLL contributions and to evaluate the impact of the threshold-resummation improved PDFs, respectively, when addressing the quark-initiated Drell-Yan production channel. Both the integrated cross section and the differential distributions with respect to the transverse momentum and invariant mass of the produced scalar-pseudoscalar pair are provided. For quark-antiquark annihilation channel, the NLO QCD relative correction can exceed $30\%$ in the low Higgs mass region, but decreases rapidly as the increment of $m_{\phi^0}$ and $m_{A^0}$. The relative correction induced by the threshold-resummation improved PDFs and the NLL resummation correction from the partonic matrix element, $K_{\text{PDF}} - 1$ and $K_{\text{PME}} - 1$, are insensitive to the mass splitting between $\phi^0$ and $A^0$, and decreases and increases respectively as the increment of Higgs mass. They could be neglected compared to the NLO QCD correction in the low invariant mass region, but become increasingly important as the increment of the invariant mass of $\phi^0A^0$, and can even predominate in the high invariant mass region. Moreover, the anomalous behavior of the NLO QCD corrected transverse momentum distribution in the small-$p_T$ region can be resolved, and the scale uncertainty can be heavily reduced, especially in the intermediate-$p_T$ region, by including the NLL resummation correction. Compared to the quark-initiated Drell-Yan channel, the contribution from the gluon-gluon fusion channel is negligible in the high invariant mass region, but indispensable and even comparable to the NLO QCD correction near the production threshold.

\vskip 5mm

\noindent{\large\bf Acknowledgments:}

This work is supported in part by the National Natural Science Foundation of China (Grants No. 11775211, No. 12061141005, No. 11805001 and No. 11935001) and the CAS Center for Excellence in Particle Physics (CCEPP).

\nocite{*}
\bibliography{ref}

%apsrev4-2.bst 2019-01-14 (MD) hand-edited version of apsrev4-1.bst
%Control: key (0)
%Control: author (8) initials jnrlst
%Control: editor formatted (1) identically to author
%Control: production of article title (0) allowed
%Control: page (0) single
%Control: year (1) truncated
%Control: production of eprint (0) enabled
\begin{thebibliography}{50}%
\makeatletter
\providecommand \@ifxundefined [1]{%
 \@ifx{#1\undefined}
}%
\providecommand \@ifnum [1]{%
 \ifnum #1\expandafter \@firstoftwo
 \else \expandafter \@secondoftwo
 \fi
}%
\providecommand \@ifx [1]{%
 \ifx #1\expandafter \@firstoftwo
 \else \expandafter \@secondoftwo
 \fi
}%
\providecommand \natexlab [1]{#1}%
\providecommand \enquote  [1]{``#1''}%
\providecommand \bibnamefont  [1]{#1}%
\providecommand \bibfnamefont [1]{#1}%
\providecommand \citenamefont [1]{#1}%
\providecommand \href@noop [0]{\@secondoftwo}%
\providecommand \href [0]{\begingroup \@sanitize@url \@href}%
\providecommand \@href[1]{\@@startlink{#1}\@@href}%
\providecommand \@@href[1]{\endgroup#1\@@endlink}%
\providecommand \@sanitize@url [0]{\catcode `\\12\catcode `\$12\catcode
  `\&12\catcode `\#12\catcode `\^12\catcode `\_12\catcode `\%12\relax}%
\providecommand \@@startlink[1]{}%
\providecommand \@@endlink[0]{}%
\providecommand \url  [0]{\begingroup\@sanitize@url \@url }%
\providecommand \@url [1]{\endgroup\@href {#1}{\urlprefix }}%
\providecommand \urlprefix  [0]{URL }%
\providecommand \Eprint [0]{\href }%
\providecommand \doibase [0]{https://doi.org/}%
\providecommand \selectlanguage [0]{\@gobble}%
\providecommand \bibinfo  [0]{\@secondoftwo}%
\providecommand \bibfield  [0]{\@secondoftwo}%
\providecommand \translation [1]{[#1]}%
\providecommand \BibitemOpen [0]{}%
\providecommand \bibitemStop [0]{}%
\providecommand \bibitemNoStop [0]{.\EOS\space}%
\providecommand \EOS [0]{\spacefactor3000\relax}%
\providecommand \BibitemShut  [1]{\csname bibitem#1\endcsname}%
\let\auto@bib@innerbib\@empty
%</preamble>
\bibitem [{\citenamefont {Aad}\ \emph {et~al.}(2012)\citenamefont {Aad} \emph
  {et~al.}}]{Aad:2012tfa}%
  \BibitemOpen
  \bibfield  {author} {\bibinfo {author} {\bibfnamefont {G.}~\bibnamefont
  {Aad}} \emph {et~al.} (\bibinfo {collaboration} {ATLAS Collaboration}),\
  }\bibfield  {title} {\bibinfo {title} {{Observation of a new particle in the
  search for the Standard Model Higgs boson with the ATLAS detector at the
  LHC}},\ }\href {https://doi.org/10.1016/j.physletb.2012.08.020} {\bibfield
  {journal} {\bibinfo  {journal} {Phys. Lett. B}\ }\textbf {\bibinfo {volume}
  {716}},\ \bibinfo {pages} {1} (\bibinfo {year} {2012})},\ \Eprint
  {https://arxiv.org/abs/1207.7214} {arXiv:1207.7214 [hep-ex]} \BibitemShut
  {NoStop}%
\bibitem [{\citenamefont {Chatrchyan}\ \emph {et~al.}(2012)\citenamefont
  {Chatrchyan} \emph {et~al.}}]{Chatrchyan:2012xdj}%
  \BibitemOpen
  \bibfield  {author} {\bibinfo {author} {\bibfnamefont {S.}~\bibnamefont
  {Chatrchyan}} \emph {et~al.} (\bibinfo {collaboration} {CMS Collaboration}),\
  }\bibfield  {title} {\bibinfo {title} {{Observation of a new boson at a mass
  of $125~ \text{GeV}$ with the CMS experiment at the LHC}},\ }\href
  {https://doi.org/10.1016/j.physletb.2012.08.021} {\bibfield  {journal}
  {\bibinfo  {journal} {Phys. Lett. B}\ }\textbf {\bibinfo {volume} {716}},\
  \bibinfo {pages} {30} (\bibinfo {year} {2012})},\ \Eprint
  {https://arxiv.org/abs/1207.7235} {arXiv:1207.7235 [hep-ex]} \BibitemShut
  {NoStop}%
\bibitem [{\citenamefont {Branco}\ \emph {et~al.}(2012)\citenamefont {Branco},
  \citenamefont {Ferreira}, \citenamefont {Lavoura}, \citenamefont {Rebelo},
  \citenamefont {Sher},\ and\ \citenamefont {Silva}}]{Branco:2011iw}%
  \BibitemOpen
  \bibfield  {author} {\bibinfo {author} {\bibfnamefont {G.~C.}\ \bibnamefont
  {Branco}}, \bibinfo {author} {\bibfnamefont {P.~M.}\ \bibnamefont
  {Ferreira}}, \bibinfo {author} {\bibfnamefont {L.}~\bibnamefont {Lavoura}},
  \bibinfo {author} {\bibfnamefont {M.~N.}\ \bibnamefont {Rebelo}}, \bibinfo
  {author} {\bibfnamefont {M.}~\bibnamefont {Sher}},\ and\ \bibinfo {author}
  {\bibfnamefont {J.~P.}\ \bibnamefont {Silva}},\ }\bibfield  {title} {\bibinfo
  {title} {{Theory and phenomenology of two-Higgs-doublet models}},\ }\href
  {https://doi.org/10.1016/j.physrep.2012.02.002} {\bibfield  {journal}
  {\bibinfo  {journal} {Phys. Rept.}\ }\textbf {\bibinfo {volume} {516}},\
  \bibinfo {pages} {1} (\bibinfo {year} {2012})},\ \Eprint
  {https://arxiv.org/abs/1106.0034} {arXiv:1106.0034 [hep-ph]} \BibitemShut
  {NoStop}%
\bibitem [{\citenamefont {Hespel}\ \emph {et~al.}(2014)\citenamefont {Hespel},
  \citenamefont {L{\' o}pez-Val},\ and\ \citenamefont
  {Vryonidou}}]{Hespel:2014sla}%
  \BibitemOpen
  \bibfield  {author} {\bibinfo {author} {\bibfnamefont {B.}~\bibnamefont
  {Hespel}}, \bibinfo {author} {\bibfnamefont {D.}~\bibnamefont {L{\'
  o}pez-Val}},\ and\ \bibinfo {author} {\bibfnamefont {E.}~\bibnamefont
  {Vryonidou}},\ }\bibfield  {title} {\bibinfo {title} {{Higgs pair production
  via gluon fusion in the Two-Higgs-Doublet Model}},\ }\href
  {https://doi.org/10.1007/JHEP09(2014)124} {\bibfield  {journal} {\bibinfo
  {journal} {JHEP}\ }\textbf {\bibinfo {volume} {09}},\ \bibinfo {pages}
  {124}},\ \Eprint {https://arxiv.org/abs/1407.0281} {arXiv:1407.0281 [hep-ph]}
  \BibitemShut {NoStop}%
\bibitem [{\citenamefont {Enberg}\ \emph {et~al.}(2017)\citenamefont {Enberg},
  \citenamefont {Klemm}, \citenamefont {Moretti},\ and\ \citenamefont
  {Munir}}]{Enberg:2016ygw}%
  \BibitemOpen
  \bibfield  {author} {\bibinfo {author} {\bibfnamefont {R.}~\bibnamefont
  {Enberg}}, \bibinfo {author} {\bibfnamefont {W.}~\bibnamefont {Klemm}},
  \bibinfo {author} {\bibfnamefont {S.}~\bibnamefont {Moretti}},\ and\ \bibinfo
  {author} {\bibfnamefont {S.}~\bibnamefont {Munir}},\ }\bibfield  {title}
  {\bibinfo {title} {{Electroweak production of light
  scalar\textendash{pseudoscalar} pairs from extended Higgs sectors}},\ }\href
  {https://doi.org/10.1016/j.physletb.2016.11.012} {\bibfield  {journal}
  {\bibinfo  {journal} {Phys. Lett. B}\ }\textbf {\bibinfo {volume} {764}},\
  \bibinfo {pages} {121} (\bibinfo {year} {2017})},\ \Eprint
  {https://arxiv.org/abs/1605.02498} {arXiv:1605.02498 [hep-ph]} \BibitemShut
  {NoStop}%
\bibitem [{\citenamefont {Sirunyan}\ \emph {et~al.}(2020)\citenamefont
  {Sirunyan} \emph {et~al.}}]{Sirunyan:2019wrn}%
  \BibitemOpen
  \bibfield  {author} {\bibinfo {author} {\bibfnamefont {A.~M.}\ \bibnamefont
  {Sirunyan}} \emph {et~al.} (\bibinfo {collaboration} {CMS Collaboration}),\
  }\bibfield  {title} {\bibinfo {title} {{Search for new neutral Higgs bosons
  through the $H \rightarrow ZA \rightarrow \ell^{+}\ell^{-} b\bar{b}$ process
  in $pp$ collisions at $\sqrt{s} = 13~ \text{TeV}$}},\ }\href
  {https://doi.org/10.1007/JHEP03(2020)055} {\bibfield  {journal} {\bibinfo
  {journal} {JHEP}\ }\textbf {\bibinfo {volume} {03}},\ \bibinfo {pages}
  {055}},\ \Eprint {https://arxiv.org/abs/1911.03781} {arXiv:1911.03781
  [hep-ex]} \BibitemShut {NoStop}%
\bibitem [{\citenamefont {Aad}\ \emph {et~al.}(2021)\citenamefont {Aad} \emph
  {et~al.}}]{Aad:2020ncx}%
  \BibitemOpen
  \bibfield  {author} {\bibinfo {author} {\bibfnamefont {G.}~\bibnamefont
  {Aad}} \emph {et~al.} (\bibinfo {collaboration} {ATLAS Collaboration}),\
  }\bibfield  {title} {\bibinfo {title} {{Search for a heavy Higgs boson
  decaying into a $Z$ boson and another heavy Higgs boson in the $\ell\ell bb$
  and $\ell\ell WW$ final states in $pp$ collisions at $\sqrt{s} = 13~
  \text{TeV}$ with the ATLAS detector}},\ }\href
  {https://doi.org/10.1140/epjc/s10052-021-09117-5} {\bibfield  {journal}
  {\bibinfo  {journal} {Eur. Phys. J. C}\ }\textbf {\bibinfo {volume} {81}},\
  \bibinfo {pages} {396} (\bibinfo {year} {2021})},\ \Eprint
  {https://arxiv.org/abs/2011.05639} {arXiv:2011.05639 [hep-ex]} \BibitemShut
  {NoStop}%
\bibitem [{\citenamefont {Dawson}\ \emph {et~al.}(1998)\citenamefont {Dawson},
  \citenamefont {Dittmaier},\ and\ \citenamefont {Spira}}]{Dawson:1998py}%
  \BibitemOpen
  \bibfield  {author} {\bibinfo {author} {\bibfnamefont {S.}~\bibnamefont
  {Dawson}}, \bibinfo {author} {\bibfnamefont {S.}~\bibnamefont {Dittmaier}},\
  and\ \bibinfo {author} {\bibfnamefont {M.}~\bibnamefont {Spira}},\ }\bibfield
   {title} {\bibinfo {title} {{Neutral Higgs-boson pair production at hadron
  colliders: QCD corrections}},\ }\href
  {https://doi.org/10.1103/PhysRevD.58.115012} {\bibfield  {journal} {\bibinfo
  {journal} {Phys. Rev. D}\ }\textbf {\bibinfo {volume} {58}},\ \bibinfo
  {pages} {115012} (\bibinfo {year} {1998})},\ \Eprint
  {https://arxiv.org/abs/hep-ph/9805244} {arXiv:hep-ph/9805244 [hep-ph]}
  \BibitemShut {NoStop}%
\bibitem [{\citenamefont {Collins}\ and\ \citenamefont
  {Soper}(1982)}]{Collins:1981va}%
  \BibitemOpen
  \bibfield  {author} {\bibinfo {author} {\bibfnamefont {J.~C.}\ \bibnamefont
  {Collins}}\ and\ \bibinfo {author} {\bibfnamefont {D.~E.}\ \bibnamefont
  {Soper}},\ }\bibfield  {title} {\bibinfo {title} {{Back-to-back jets: Fourier
  transform from $b$ to $k_T$}},\ }\href
  {https://doi.org/10.1016/0550-3213(82)90453-9} {\bibfield  {journal}
  {\bibinfo  {journal} {Nucl. Phys. B}\ }\textbf {\bibinfo {volume} {197}},\
  \bibinfo {pages} {446} (\bibinfo {year} {1982})}\BibitemShut {NoStop}%
\bibitem [{\citenamefont {Collins}\ and\ \citenamefont
  {Soper}(1981)}]{Collins:1981uk}%
  \BibitemOpen
  \bibfield  {author} {\bibinfo {author} {\bibfnamefont {J.~C.}\ \bibnamefont
  {Collins}}\ and\ \bibinfo {author} {\bibfnamefont {D.~E.}\ \bibnamefont
  {Soper}},\ }\bibfield  {title} {\bibinfo {title} {{Back-to-back jets in
  QCD}},\ }\href {https://doi.org/10.1016/0550-3213(81)90339-4} {\bibfield
  {journal} {\bibinfo  {journal} {Nucl. Phys. B}\ }\textbf {\bibinfo {volume}
  {193}},\ \bibinfo {pages} {381} (\bibinfo {year} {1981})},\ \bibinfo {note}
  {[Erratum: Nucl. Phys. B 213, 545 (1983)]}\BibitemShut {NoStop}%
\bibitem [{\citenamefont {Collins}\ \emph {et~al.}(1985)\citenamefont
  {Collins}, \citenamefont {Soper},\ and\ \citenamefont
  {Sterman}}]{Collins:1984kg}%
  \BibitemOpen
  \bibfield  {author} {\bibinfo {author} {\bibfnamefont {J.~C.}\ \bibnamefont
  {Collins}}, \bibinfo {author} {\bibfnamefont {D.~E.}\ \bibnamefont {Soper}},\
  and\ \bibinfo {author} {\bibfnamefont {G.}~\bibnamefont {Sterman}},\
  }\bibfield  {title} {\bibinfo {title} {{Transverse momentum distribution in
  Drell-Yan pair and $W$ and $Z$ boson production}},\ }\href
  {https://doi.org/10.1016/0550-3213(85)90479-1} {\bibfield  {journal}
  {\bibinfo  {journal} {Nucl. Phys. B}\ }\textbf {\bibinfo {volume} {250}},\
  \bibinfo {pages} {199} (\bibinfo {year} {1985})}\BibitemShut {NoStop}%
\bibitem [{\citenamefont {Sterman}(1987)}]{Sterman:1986aj}%
  \BibitemOpen
  \bibfield  {author} {\bibinfo {author} {\bibfnamefont {G.}~\bibnamefont
  {Sterman}},\ }\bibfield  {title} {\bibinfo {title} {{Summation of large
  corrections to short-distance hadronic cross sections}},\ }\href
  {https://doi.org/10.1016/0550-3213(87)90258-6} {\bibfield  {journal}
  {\bibinfo  {journal} {Nucl. Phys. B}\ }\textbf {\bibinfo {volume} {281}},\
  \bibinfo {pages} {310} (\bibinfo {year} {1987})}\BibitemShut {NoStop}%
\bibitem [{\citenamefont {Catani}\ and\ \citenamefont
  {Trentadue}(1989)}]{Catani:1989ne}%
  \BibitemOpen
  \bibfield  {author} {\bibinfo {author} {\bibfnamefont {S.}~\bibnamefont
  {Catani}}\ and\ \bibinfo {author} {\bibfnamefont {L.}~\bibnamefont
  {Trentadue}},\ }\bibfield  {title} {\bibinfo {title} {{Resummation of the QCD
  perturbative series for hard processes}},\ }\href
  {https://doi.org/10.1016/0550-3213(89)90273-3} {\bibfield  {journal}
  {\bibinfo  {journal} {Nucl. Phys. B}\ }\textbf {\bibinfo {volume} {327}},\
  \bibinfo {pages} {323} (\bibinfo {year} {1989})}\BibitemShut {NoStop}%
\bibitem [{\citenamefont {Catani}\ \emph {et~al.}(1996)\citenamefont {Catani},
  \citenamefont {Mangano}, \citenamefont {Nason},\ and\ \citenamefont
  {Trentadue}}]{Catani:1996yz}%
  \BibitemOpen
  \bibfield  {author} {\bibinfo {author} {\bibfnamefont {S.}~\bibnamefont
  {Catani}}, \bibinfo {author} {\bibfnamefont {M.~L.}\ \bibnamefont {Mangano}},
  \bibinfo {author} {\bibfnamefont {P.}~\bibnamefont {Nason}},\ and\ \bibinfo
  {author} {\bibfnamefont {L.}~\bibnamefont {Trentadue}},\ }\bibfield  {title}
  {\bibinfo {title} {{The resummation of soft gluons in hadronic collisions}},\
  }\href {https://doi.org/10.1016/0550-3213(96)00399-9} {\bibfield  {journal}
  {\bibinfo  {journal} {Nucl. Phys. B}\ }\textbf {\bibinfo {volume} {478}},\
  \bibinfo {pages} {273} (\bibinfo {year} {1996})},\ \Eprint
  {https://arxiv.org/abs/hep-ph/9604351} {arXiv:hep-ph/9604351 [hep-ph]}
  \BibitemShut {NoStop}%
\bibitem [{\citenamefont {Catani}\ \emph
  {et~al.}(2001{\natexlab{a}})\citenamefont {Catani}, \citenamefont
  {de~Florian},\ and\ \citenamefont {Grazzini}}]{Catani:2000vq}%
  \BibitemOpen
  \bibfield  {author} {\bibinfo {author} {\bibfnamefont {S.}~\bibnamefont
  {Catani}}, \bibinfo {author} {\bibfnamefont {D.}~\bibnamefont {de~Florian}},\
  and\ \bibinfo {author} {\bibfnamefont {M.}~\bibnamefont {Grazzini}},\
  }\bibfield  {title} {\bibinfo {title} {{Universality of non-leading
  logarithmic contributions in transverse-momentum distributions}},\ }\href
  {https://doi.org/10.1016/S0550-3213(00)00617-9} {\bibfield  {journal}
  {\bibinfo  {journal} {Nucl. Phys. B}\ }\textbf {\bibinfo {volume} {596}},\
  \bibinfo {pages} {299} (\bibinfo {year} {2001}{\natexlab{a}})},\ \Eprint
  {https://arxiv.org/abs/hep-ph/0008184} {arXiv:hep-ph/0008184 [hep-ph]}
  \BibitemShut {NoStop}%
\bibitem [{\citenamefont {Bozzi}\ \emph {et~al.}(2006)\citenamefont {Bozzi},
  \citenamefont {Catani}, \citenamefont {de~Florian},\ and\ \citenamefont
  {Grazzini}}]{Bozzi:2005wk}%
  \BibitemOpen
  \bibfield  {author} {\bibinfo {author} {\bibfnamefont {G.}~\bibnamefont
  {Bozzi}}, \bibinfo {author} {\bibfnamefont {S.}~\bibnamefont {Catani}},
  \bibinfo {author} {\bibfnamefont {D.}~\bibnamefont {de~Florian}},\ and\
  \bibinfo {author} {\bibfnamefont {M.}~\bibnamefont {Grazzini}},\ }\bibfield
  {title} {\bibinfo {title} {{Transverse-momentum resummation and the spectrum
  of the Higgs boson at the LHC}},\ }\href
  {https://doi.org/10.1016/j.nuclphysb.2005.12.022} {\bibfield  {journal}
  {\bibinfo  {journal} {Nucl. Phys. B}\ }\textbf {\bibinfo {volume} {737}},\
  \bibinfo {pages} {73} (\bibinfo {year} {2006})},\ \Eprint
  {https://arxiv.org/abs/hep-ph/0508068} {arXiv:hep-ph/0508068 [hep-ph]}
  \BibitemShut {NoStop}%
\bibitem [{\citenamefont {Beenakker}\ \emph {et~al.}(2016)\citenamefont
  {Beenakker}, \citenamefont {Borschensky}, \citenamefont {Kr{\" a}mer},
  \citenamefont {Kulesza}, \citenamefont {Laenen}, \citenamefont {Marzani},\
  and\ \citenamefont {Rojo}}]{Beenakker:2015rna}%
  \BibitemOpen
  \bibfield  {author} {\bibinfo {author} {\bibfnamefont {W.}~\bibnamefont
  {Beenakker}}, \bibinfo {author} {\bibfnamefont {C.}~\bibnamefont
  {Borschensky}}, \bibinfo {author} {\bibfnamefont {M.}~\bibnamefont {Kr{\"
  a}mer}}, \bibinfo {author} {\bibfnamefont {A.}~\bibnamefont {Kulesza}},
  \bibinfo {author} {\bibfnamefont {E.}~\bibnamefont {Laenen}}, \bibinfo
  {author} {\bibfnamefont {S.}~\bibnamefont {Marzani}},\ and\ \bibinfo {author}
  {\bibfnamefont {J.}~\bibnamefont {Rojo}},\ }\bibfield  {title} {\bibinfo
  {title} {{NLO+NLL squark and gluino production cross sections with
  threshold-improved parton distributions}},\ }\href
  {https://doi.org/10.1140/epjc/s10052-016-3892-4} {\bibfield  {journal}
  {\bibinfo  {journal} {Eur. Phys. J. C}\ }\textbf {\bibinfo {volume} {76}},\
  \bibinfo {pages} {53} (\bibinfo {year} {2016})},\ \Eprint
  {https://arxiv.org/abs/1510.00375} {arXiv:1510.00375 [hep-ph]} \BibitemShut
  {NoStop}%
\bibitem [{\citenamefont {Eriksson}\ \emph {et~al.}(2010)\citenamefont
  {Eriksson}, \citenamefont {Rathsman},\ and\ \citenamefont {St{\r
  a}l}}]{Eriksson:2010zzb}%
  \BibitemOpen
  \bibfield  {author} {\bibinfo {author} {\bibfnamefont {D.}~\bibnamefont
  {Eriksson}}, \bibinfo {author} {\bibfnamefont {J.}~\bibnamefont {Rathsman}},\
  and\ \bibinfo {author} {\bibfnamefont {O.}~\bibnamefont {St{\r a}l}},\
  }\bibfield  {title} {\bibinfo {title} {{2HDMC $-$ two-Higgs-doublet model
  calculator}},\ }\href {https://doi.org/10.1016/j.cpc.2009.12.016} {\bibfield
  {journal} {\bibinfo  {journal} {Comput. Phys. Commun.}\ }\textbf {\bibinfo
  {volume} {181}},\ \bibinfo {pages} {833} (\bibinfo {year} {2010})},\ \Eprint
  {https://arxiv.org/abs/0902.0851} {arXiv:0902.0851 [hep-ph]} \BibitemShut
  {NoStop}%
\bibitem [{\citenamefont {'t~Hooft}\ and\ \citenamefont
  {Veltman}(1972)}]{tHooft:1972tcz}%
  \BibitemOpen
  \bibfield  {author} {\bibinfo {author} {\bibfnamefont {G.}~\bibnamefont
  {'t~Hooft}}\ and\ \bibinfo {author} {\bibfnamefont {M.}~\bibnamefont
  {Veltman}},\ }\bibfield  {title} {\bibinfo {title} {{Regularization and
  renormalization of gauge fields}},\ }\href
  {https://doi.org/10.1016/0550-3213(72)90279-9} {\bibfield  {journal}
  {\bibinfo  {journal} {Nucl. Phys. B}\ }\textbf {\bibinfo {volume} {44}},\
  \bibinfo {pages} {189} (\bibinfo {year} {1972})}\BibitemShut {NoStop}%
\bibitem [{\citenamefont {Catani}\ and\ \citenamefont
  {Seymour}(1997)}]{Catani:1996vz}%
  \BibitemOpen
  \bibfield  {author} {\bibinfo {author} {\bibfnamefont {S.}~\bibnamefont
  {Catani}}\ and\ \bibinfo {author} {\bibfnamefont {M.~H.}\ \bibnamefont
  {Seymour}},\ }\bibfield  {title} {\bibinfo {title} {{A general algorithm for
  calculating jet cross sections in NLO QCD}},\ }\href
  {https://doi.org/10.1016/S0550-3213(96)00589-5,
  10.1016/S0550-3213(98)81022-5} {\bibfield  {journal} {\bibinfo  {journal}
  {Nucl. Phys. B}\ }\textbf {\bibinfo {volume} {485}},\ \bibinfo {pages} {291}
  (\bibinfo {year} {1997})},\ \bibinfo {note} {[Erratum: Nucl. Phys. B 510, 503
  (1998)]},\ \Eprint {https://arxiv.org/abs/hep-ph/9605323}
  {arXiv:hep-ph/9605323 [hep-ph]} \BibitemShut {NoStop}%
\bibitem [{\citenamefont {Harris}\ and\ \citenamefont
  {Owens}(2002)}]{Harris:2001sx}%
  \BibitemOpen
  \bibfield  {author} {\bibinfo {author} {\bibfnamefont {B.~W.}\ \bibnamefont
  {Harris}}\ and\ \bibinfo {author} {\bibfnamefont {J.~F.}\ \bibnamefont
  {Owens}},\ }\bibfield  {title} {\bibinfo {title} {{Two cutoff phase space
  slicing method}},\ }\href {https://doi.org/10.1103/PhysRevD.65.094032}
  {\bibfield  {journal} {\bibinfo  {journal} {Phys. Rev. D}\ }\textbf {\bibinfo
  {volume} {65}},\ \bibinfo {pages} {094032} (\bibinfo {year} {2002})},\
  \Eprint {https://arxiv.org/abs/hep-ph/0102128} {arXiv:hep-ph/0102128
  [hep-ph]} \BibitemShut {NoStop}%
\bibitem [{\citenamefont {Dulat}\ \emph {et~al.}(2016)\citenamefont {Dulat},
  \citenamefont {Hou}, \citenamefont {Gao}, \citenamefont {Guzzi},
  \citenamefont {Huston}, \citenamefont {Nadolsky}, \citenamefont {Pumplin},
  \citenamefont {Schmidt}, \citenamefont {Stump},\ and\ \citenamefont
  {Yuan}}]{Dulat:2015mca}%
  \BibitemOpen
  \bibfield  {author} {\bibinfo {author} {\bibfnamefont {S.}~\bibnamefont
  {Dulat}}, \bibinfo {author} {\bibfnamefont {T.-J.}\ \bibnamefont {Hou}},
  \bibinfo {author} {\bibfnamefont {J.}~\bibnamefont {Gao}}, \bibinfo {author}
  {\bibfnamefont {M.}~\bibnamefont {Guzzi}}, \bibinfo {author} {\bibfnamefont
  {J.}~\bibnamefont {Huston}}, \bibinfo {author} {\bibfnamefont
  {P.}~\bibnamefont {Nadolsky}}, \bibinfo {author} {\bibfnamefont
  {J.}~\bibnamefont {Pumplin}}, \bibinfo {author} {\bibfnamefont
  {C.}~\bibnamefont {Schmidt}}, \bibinfo {author} {\bibfnamefont
  {D.}~\bibnamefont {Stump}},\ and\ \bibinfo {author} {\bibfnamefont {C.-P.}\
  \bibnamefont {Yuan}},\ }\bibfield  {title} {\bibinfo {title} {{New parton
  distribution functions from a global analysis of quantum chromodynamics}},\
  }\href {https://doi.org/10.1103/PhysRevD.93.033006} {\bibfield  {journal}
  {\bibinfo  {journal} {Phys. Rev. D}\ }\textbf {\bibinfo {volume} {93}},\
  \bibinfo {pages} {033006} (\bibinfo {year} {2016})},\ \Eprint
  {https://arxiv.org/abs/1506.07443} {arXiv:1506.07443 [hep-ph]} \BibitemShut
  {NoStop}%
\bibitem [{\citenamefont {Contopanagos}\ \emph {et~al.}(1997)\citenamefont
  {Contopanagos}, \citenamefont {Laenen},\ and\ \citenamefont
  {Sterman}}]{Contopanagos:1996nh}%
  \BibitemOpen
  \bibfield  {author} {\bibinfo {author} {\bibfnamefont {H.}~\bibnamefont
  {Contopanagos}}, \bibinfo {author} {\bibfnamefont {E.}~\bibnamefont
  {Laenen}},\ and\ \bibinfo {author} {\bibfnamefont {G.}~\bibnamefont
  {Sterman}},\ }\bibfield  {title} {\bibinfo {title} {{Sudakov factorization
  and resummation}},\ }\href {https://doi.org/10.1016/S0550-3213(96)00567-6}
  {\bibfield  {journal} {\bibinfo  {journal} {Nucl. Phys. B}\ }\textbf
  {\bibinfo {volume} {484}},\ \bibinfo {pages} {303} (\bibinfo {year}
  {1997})},\ \Eprint {https://arxiv.org/abs/hep-ph/9604313}
  {arXiv:hep-ph/9604313 [hep-ph]} \BibitemShut {NoStop}%
\bibitem [{\citenamefont {Catani}\ and\ \citenamefont
  {Trentadue}(1991)}]{Catani:1990rp}%
  \BibitemOpen
  \bibfield  {author} {\bibinfo {author} {\bibfnamefont {S.}~\bibnamefont
  {Catani}}\ and\ \bibinfo {author} {\bibfnamefont {L.}~\bibnamefont
  {Trentadue}},\ }\bibfield  {title} {\bibinfo {title} {{Comment on QCD
  exponentiation at large $x$}},\ }\href
  {https://doi.org/10.1016/0550-3213(91)90506-S} {\bibfield  {journal}
  {\bibinfo  {journal} {Nucl. Phys. B}\ }\textbf {\bibinfo {volume} {353}},\
  \bibinfo {pages} {183} (\bibinfo {year} {1991})}\BibitemShut {NoStop}%
\bibitem [{\citenamefont {Kidonakis}\ and\ \citenamefont
  {Sterman}(1997)}]{Kidonakis:1997gm}%
  \BibitemOpen
  \bibfield  {author} {\bibinfo {author} {\bibfnamefont {N.}~\bibnamefont
  {Kidonakis}}\ and\ \bibinfo {author} {\bibfnamefont {G.}~\bibnamefont
  {Sterman}},\ }\bibfield  {title} {\bibinfo {title} {{Resummation for QCD hard
  scattering}},\ }\href {https://doi.org/10.1016/S0550-3213(97)00506-3}
  {\bibfield  {journal} {\bibinfo  {journal} {Nucl. Phys. B}\ }\textbf
  {\bibinfo {volume} {505}},\ \bibinfo {pages} {321} (\bibinfo {year}
  {1997})},\ \Eprint {https://arxiv.org/abs/hep-ph/9705234}
  {arXiv:hep-ph/9705234 [hep-ph]} \BibitemShut {NoStop}%
\bibitem [{\citenamefont {Kidonakis}\ \emph {et~al.}(1998)\citenamefont
  {Kidonakis}, \citenamefont {Oderda},\ and\ \citenamefont
  {Sterman}}]{Kidonakis:1998bk}%
  \BibitemOpen
  \bibfield  {author} {\bibinfo {author} {\bibfnamefont {N.}~\bibnamefont
  {Kidonakis}}, \bibinfo {author} {\bibfnamefont {G.}~\bibnamefont {Oderda}},\
  and\ \bibinfo {author} {\bibfnamefont {G.}~\bibnamefont {Sterman}},\
  }\bibfield  {title} {\bibinfo {title} {{Threshold resummation for dijet cross
  sections}},\ }\href {https://doi.org/10.1016/S0550-3213(98)00243-0}
  {\bibfield  {journal} {\bibinfo  {journal} {Nucl. Phys. B}\ }\textbf
  {\bibinfo {volume} {525}},\ \bibinfo {pages} {299} (\bibinfo {year}
  {1998})},\ \Eprint {https://arxiv.org/abs/hep-ph/9801268}
  {arXiv:hep-ph/9801268 [hep-ph]} \BibitemShut {NoStop}%
\bibitem [{\citenamefont {Vogt}(2001)}]{Vogt:2000ci}%
  \BibitemOpen
  \bibfield  {author} {\bibinfo {author} {\bibfnamefont {A.}~\bibnamefont
  {Vogt}},\ }\bibfield  {title} {\bibinfo {title} {{Next-to-next-to-leading
  logarithmic threshold resummation for deep-inelastic scattering and the
  Drell-Yan process}},\ }\href {https://doi.org/10.1016/S0370-2693(00)01344-7}
  {\bibfield  {journal} {\bibinfo  {journal} {Phys. Lett. B}\ }\textbf
  {\bibinfo {volume} {497}},\ \bibinfo {pages} {228} (\bibinfo {year}
  {2001})},\ \Eprint {https://arxiv.org/abs/hep-ph/0010146}
  {arXiv:hep-ph/0010146 [hep-ph]} \BibitemShut {NoStop}%
\bibitem [{\citenamefont {Kr{\" a}mer}\ \emph {et~al.}(1998)\citenamefont
  {Kr{\" a}mer}, \citenamefont {Laenen},\ and\ \citenamefont
  {Spira}}]{Kramer:1996iq}%
  \BibitemOpen
  \bibfield  {author} {\bibinfo {author} {\bibfnamefont {M.}~\bibnamefont
  {Kr{\" a}mer}}, \bibinfo {author} {\bibfnamefont {E.}~\bibnamefont
  {Laenen}},\ and\ \bibinfo {author} {\bibfnamefont {M.}~\bibnamefont
  {Spira}},\ }\bibfield  {title} {\bibinfo {title} {{Soft gluon radiation in
  Higgs boson production at the LHC}},\ }\href
  {https://doi.org/10.1016/S0550-3213(97)00679-2} {\bibfield  {journal}
  {\bibinfo  {journal} {Nucl. Phys. B}\ }\textbf {\bibinfo {volume} {511}},\
  \bibinfo {pages} {523} (\bibinfo {year} {1998})},\ \Eprint
  {https://arxiv.org/abs/hep-ph/9611272} {arXiv:hep-ph/9611272 [hep-ph]}
  \BibitemShut {NoStop}%
\bibitem [{\citenamefont {Catani}\ \emph
  {et~al.}(2001{\natexlab{b}})\citenamefont {Catani}, \citenamefont
  {de~Florian},\ and\ \citenamefont {Grazzini}}]{Catani:2001ic}%
  \BibitemOpen
  \bibfield  {author} {\bibinfo {author} {\bibfnamefont {S.}~\bibnamefont
  {Catani}}, \bibinfo {author} {\bibfnamefont {D.}~\bibnamefont {de~Florian}},\
  and\ \bibinfo {author} {\bibfnamefont {M.}~\bibnamefont {Grazzini}},\
  }\bibfield  {title} {\bibinfo {title} {{Higgs production in hadron
  collisions: soft and virtual QCD corrections at NNLO}},\ }\href
  {https://doi.org/10.1088/1126-6708/2001/05/025} {\bibfield  {journal}
  {\bibinfo  {journal} {JHEP}\ }\textbf {\bibinfo {volume} {05}},\ \bibinfo
  {pages} {025}},\ \Eprint {https://arxiv.org/abs/hep-ph/0102227}
  {arXiv:hep-ph/0102227 [hep-ph]} \BibitemShut {NoStop}%
\bibitem [{\citenamefont {Kulesza}\ \emph {et~al.}(2002)\citenamefont
  {Kulesza}, \citenamefont {Sterman},\ and\ \citenamefont
  {Vogelsang}}]{Kulesza:2002rh}%
  \BibitemOpen
  \bibfield  {author} {\bibinfo {author} {\bibfnamefont {A.}~\bibnamefont
  {Kulesza}}, \bibinfo {author} {\bibfnamefont {G.}~\bibnamefont {Sterman}},\
  and\ \bibinfo {author} {\bibfnamefont {W.}~\bibnamefont {Vogelsang}},\
  }\bibfield  {title} {\bibinfo {title} {{Joint resummation in electroweak
  boson production}},\ }\href {https://doi.org/10.1103/PhysRevD.66.014011}
  {\bibfield  {journal} {\bibinfo  {journal} {Phys. Rev. D}\ }\textbf {\bibinfo
  {volume} {66}},\ \bibinfo {pages} {014011} (\bibinfo {year} {2002})},\
  \Eprint {https://arxiv.org/abs/hep-ph/0202251} {arXiv:hep-ph/0202251
  [hep-ph]} \BibitemShut {NoStop}%
\bibitem [{\citenamefont {Bozzi}\ \emph {et~al.}(2007)\citenamefont {Bozzi},
  \citenamefont {Fuks},\ and\ \citenamefont {Klasen}}]{Bozzi:2007qr}%
  \BibitemOpen
  \bibfield  {author} {\bibinfo {author} {\bibfnamefont {G.}~\bibnamefont
  {Bozzi}}, \bibinfo {author} {\bibfnamefont {B.}~\bibnamefont {Fuks}},\ and\
  \bibinfo {author} {\bibfnamefont {M.}~\bibnamefont {Klasen}},\ }\bibfield
  {title} {\bibinfo {title} {{Threshold resummation for slepton-pair production
  at hadron colliders}},\ }\href
  {https://doi.org/10.1016/j.nuclphysb.2007.03.052} {\bibfield  {journal}
  {\bibinfo  {journal} {Nucl. Phys. B}\ }\textbf {\bibinfo {volume} {777}},\
  \bibinfo {pages} {157} (\bibinfo {year} {2007})},\ \Eprint
  {https://arxiv.org/abs/hep-ph/0701202} {arXiv:hep-ph/0701202 [hep-ph]}
  \BibitemShut {NoStop}%
\bibitem [{\citenamefont {Almeida}\ \emph {et~al.}(2009)\citenamefont
  {Almeida}, \citenamefont {Sterman},\ and\ \citenamefont
  {Vogelsang}}]{Almeida:2009jt}%
  \BibitemOpen
  \bibfield  {author} {\bibinfo {author} {\bibfnamefont {L.~G.}\ \bibnamefont
  {Almeida}}, \bibinfo {author} {\bibfnamefont {G.}~\bibnamefont {Sterman}},\
  and\ \bibinfo {author} {\bibfnamefont {W.}~\bibnamefont {Vogelsang}},\
  }\bibfield  {title} {\bibinfo {title} {{Threshold resummation for dihadron
  production in hadronic collisions}},\ }\href
  {https://doi.org/10.1103/PhysRevD.80.074016} {\bibfield  {journal} {\bibinfo
  {journal} {Phys. Rev. D}\ }\textbf {\bibinfo {volume} {80}},\ \bibinfo
  {pages} {074016} (\bibinfo {year} {2009})},\ \Eprint
  {https://arxiv.org/abs/0907.1234} {arXiv:0907.1234 [hep-ph]} \BibitemShut
  {NoStop}%
\bibitem [{\citenamefont {Laenen}\ \emph {et~al.}(2000)\citenamefont {Laenen},
  \citenamefont {Sterman},\ and\ \citenamefont {Vogelsang}}]{Laenen:2000de}%
  \BibitemOpen
  \bibfield  {author} {\bibinfo {author} {\bibfnamefont {E.}~\bibnamefont
  {Laenen}}, \bibinfo {author} {\bibfnamefont {G.}~\bibnamefont {Sterman}},\
  and\ \bibinfo {author} {\bibfnamefont {W.}~\bibnamefont {Vogelsang}},\
  }\bibfield  {title} {\bibinfo {title} {{Higher-order QCD corrections in
  prompt photon production}},\ }\href
  {https://doi.org/10.1103/PhysRevLett.84.4296} {\bibfield  {journal} {\bibinfo
   {journal} {Phys. Rev. Lett.}\ }\textbf {\bibinfo {volume} {84}},\ \bibinfo
  {pages} {4296} (\bibinfo {year} {2000})},\ \Eprint
  {https://arxiv.org/abs/hep-ph/0002078} {arXiv:hep-ph/0002078 [hep-ph]}
  \BibitemShut {NoStop}%
\bibitem [{\citenamefont {Furmanski}\ and\ \citenamefont
  {Petronzio}(1982)}]{Furmanski:1981cw}%
  \BibitemOpen
  \bibfield  {author} {\bibinfo {author} {\bibfnamefont {W.}~\bibnamefont
  {Furmanski}}\ and\ \bibinfo {author} {\bibfnamefont {R.}~\bibnamefont
  {Petronzio}},\ }\bibfield  {title} {\bibinfo {title} {{Lepton-hadron
  processes beyond leading order in quantum chromodynamics}},\ }\href
  {https://doi.org/10.1007/BF01578280} {\bibfield  {journal} {\bibinfo
  {journal} {Z. Phys. C}\ }\textbf {\bibinfo {volume} {11}},\ \bibinfo {pages}
  {293} (\bibinfo {year} {1982})}\BibitemShut {NoStop}%
\bibitem [{\citenamefont {Debove}\ \emph {et~al.}(2011)\citenamefont {Debove},
  \citenamefont {Fuks},\ and\ \citenamefont {Klasen}}]{Debove:2010kf}%
  \BibitemOpen
  \bibfield  {author} {\bibinfo {author} {\bibfnamefont {J.}~\bibnamefont
  {Debove}}, \bibinfo {author} {\bibfnamefont {B.}~\bibnamefont {Fuks}},\ and\
  \bibinfo {author} {\bibfnamefont {M.}~\bibnamefont {Klasen}},\ }\bibfield
  {title} {\bibinfo {title} {{Threshold resummation for gaugino pair production
  at hadron colliders}},\ }\href
  {https://doi.org/10.1016/j.nuclphysb.2010.08.016} {\bibfield  {journal}
  {\bibinfo  {journal} {Nucl. Phys. B}\ }\textbf {\bibinfo {volume} {842}},\
  \bibinfo {pages} {51} (\bibinfo {year} {2011})},\ \Eprint
  {https://arxiv.org/abs/1005.2909} {arXiv:1005.2909 [hep-ph]} \BibitemShut
  {NoStop}%
\bibitem [{\citenamefont {Fuks}\ \emph {et~al.}(2013)\citenamefont {Fuks},
  \citenamefont {Klasen}, \citenamefont {Lamprea},\ and\ \citenamefont
  {Rothering}}]{Fuks:2013vua}%
  \BibitemOpen
  \bibfield  {author} {\bibinfo {author} {\bibfnamefont {B.}~\bibnamefont
  {Fuks}}, \bibinfo {author} {\bibfnamefont {M.}~\bibnamefont {Klasen}},
  \bibinfo {author} {\bibfnamefont {D.~R.}\ \bibnamefont {Lamprea}},\ and\
  \bibinfo {author} {\bibfnamefont {M.}~\bibnamefont {Rothering}},\ }\bibfield
  {title} {\bibinfo {title} {{Precision predictions for electroweak
  superpartner production at hadron colliders with {\sc Resummino}}},\ }\href
  {https://doi.org/10.1140/epjc/s10052-013-2480-0} {\bibfield  {journal}
  {\bibinfo  {journal} {Eur. Phys. J. C}\ }\textbf {\bibinfo {volume} {73}},\
  \bibinfo {pages} {2480} (\bibinfo {year} {2013})},\ \Eprint
  {https://arxiv.org/abs/1304.0790} {arXiv:1304.0790 [hep-ph]} \BibitemShut
  {NoStop}%
\bibitem [{\citenamefont {Bonvini}\ \emph {et~al.}(2015)\citenamefont
  {Bonvini}, \citenamefont {Marzani}, \citenamefont {Rojo}, \citenamefont
  {Rottoli}, \citenamefont {Ubiali}, \citenamefont {Ball}, \citenamefont
  {Bertone}, \citenamefont {Carrazza},\ and\ \citenamefont
  {Hartland}}]{Bonvini:2015ira}%
  \BibitemOpen
  \bibfield  {author} {\bibinfo {author} {\bibfnamefont {M.}~\bibnamefont
  {Bonvini}}, \bibinfo {author} {\bibfnamefont {S.}~\bibnamefont {Marzani}},
  \bibinfo {author} {\bibfnamefont {J.}~\bibnamefont {Rojo}}, \bibinfo {author}
  {\bibfnamefont {L.}~\bibnamefont {Rottoli}}, \bibinfo {author} {\bibfnamefont
  {M.}~\bibnamefont {Ubiali}}, \bibinfo {author} {\bibfnamefont {R.~D.}\
  \bibnamefont {Ball}}, \bibinfo {author} {\bibfnamefont {V.}~\bibnamefont
  {Bertone}}, \bibinfo {author} {\bibfnamefont {S.}~\bibnamefont {Carrazza}},\
  and\ \bibinfo {author} {\bibfnamefont {N.~P.}\ \bibnamefont {Hartland}},\
  }\bibfield  {title} {\bibinfo {title} {{Parton distributions with threshold
  resummation}},\ }\href {https://doi.org/10.1007/JHEP09(2015)191} {\bibfield
  {journal} {\bibinfo  {journal} {JHEP}\ }\textbf {\bibinfo {volume} {09}},\
  \bibinfo {pages} {191}},\ \Eprint {https://arxiv.org/abs/1507.01006}
  {arXiv:1507.01006 [hep-ph]} \BibitemShut {NoStop}%
\bibitem [{\citenamefont {Fuks}\ \emph {et~al.}(2014)\citenamefont {Fuks},
  \citenamefont {Klasen}, \citenamefont {Lamprea},\ and\ \citenamefont
  {Rothering}}]{Fuks:2013lya}%
  \BibitemOpen
  \bibfield  {author} {\bibinfo {author} {\bibfnamefont {B.}~\bibnamefont
  {Fuks}}, \bibinfo {author} {\bibfnamefont {M.}~\bibnamefont {Klasen}},
  \bibinfo {author} {\bibfnamefont {D.~R.}\ \bibnamefont {Lamprea}},\ and\
  \bibinfo {author} {\bibfnamefont {M.}~\bibnamefont {Rothering}},\ }\bibfield
  {title} {\bibinfo {title} {{Revisiting slepton pair production at the Large
  Hadron Collider}},\ }\href {https://doi.org/10.1007/JHEP01(2014)168}
  {\bibfield  {journal} {\bibinfo  {journal} {JHEP}\ }\textbf {\bibinfo
  {volume} {01}},\ \bibinfo {pages} {168}},\ \Eprint
  {https://arxiv.org/abs/1310.2621} {arXiv:1310.2621 [hep-ph]} \BibitemShut
  {NoStop}%
\bibitem [{\citenamefont {Hahn}\ and\ \citenamefont {P{\'
  e}rez-Victoria}(1999)}]{Hahn:1998yk}%
  \BibitemOpen
  \bibfield  {author} {\bibinfo {author} {\bibfnamefont {T.}~\bibnamefont
  {Hahn}}\ and\ \bibinfo {author} {\bibfnamefont {M.}~\bibnamefont {P{\'
  e}rez-Victoria}},\ }\bibfield  {title} {\bibinfo {title} {{Automatized
  one-loop calculations in four and $D$ dimensions}},\ }\href
  {https://doi.org/10.1016/S0010-4655(98)00173-8} {\bibfield  {journal}
  {\bibinfo  {journal} {Comput. Phys. Commun.}\ }\textbf {\bibinfo {volume}
  {118}},\ \bibinfo {pages} {153} (\bibinfo {year} {1999})},\ \Eprint
  {https://arxiv.org/abs/hep-ph/9807565} {arXiv:hep-ph/9807565 [hep-ph]}
  \BibitemShut {NoStop}%
\bibitem [{\citenamefont {Hahn}(2001)}]{Hahn:2000kx}%
  \BibitemOpen
  \bibfield  {author} {\bibinfo {author} {\bibfnamefont {T.}~\bibnamefont
  {Hahn}},\ }\bibfield  {title} {\bibinfo {title} {{Generating Feynman diagrams
  and amplitudes with {\it FeynArts} 3}},\ }\href
  {https://doi.org/10.1016/S0010-4655(01)00290-9} {\bibfield  {journal}
  {\bibinfo  {journal} {Comput. Phys. Commun.}\ }\textbf {\bibinfo {volume}
  {140}},\ \bibinfo {pages} {418} (\bibinfo {year} {2001})},\ \Eprint
  {https://arxiv.org/abs/hep-ph/0012260} {arXiv:hep-ph/0012260 [hep-ph]}
  \BibitemShut {NoStop}%
\bibitem [{\citenamefont {van Oldenborgh}(1991)}]{vanOldenborgh:1990yc}%
  \BibitemOpen
  \bibfield  {author} {\bibinfo {author} {\bibfnamefont {G.~J.}\ \bibnamefont
  {van Oldenborgh}},\ }\bibfield  {title} {\bibinfo {title} {{FF $-$ a package
  to evaluate one-loop Feynman diagrams}},\ }\href
  {https://doi.org/10.1016/0010-4655(91)90002-3} {\bibfield  {journal}
  {\bibinfo  {journal} {Comput. Phys. Commun.}\ }\textbf {\bibinfo {volume}
  {66}},\ \bibinfo {pages} {1} (\bibinfo {year} {1991})}\BibitemShut {NoStop}%
\bibitem [{\citenamefont {Zyla}\ \emph {et~al.}(2020)\citenamefont {Zyla} \emph
  {et~al.}}]{Zyla:2020zbs}%
  \BibitemOpen
  \bibfield  {author} {\bibinfo {author} {\bibfnamefont {P.~A.}\ \bibnamefont
  {Zyla}} \emph {et~al.} (\bibinfo {collaboration} {Particle Data Group}),\
  }\bibfield  {title} {\bibinfo {title} {{Review of particle physics}},\ }\href
  {https://doi.org/10.1093/ptep/ptaa104} {\bibfield  {journal} {\bibinfo
  {journal} {Prog. Theor. Exp. Phys.}\ }\textbf {\bibinfo {volume} {2020}},\
  \bibinfo {pages} {083C01} (\bibinfo {year} {2020})}\BibitemShut {NoStop}%
\bibitem [{\citenamefont {Grinstein}\ \emph {et~al.}(2016)\citenamefont
  {Grinstein}, \citenamefont {Murphy},\ and\ \citenamefont
  {Uttayarat}}]{Grinstein:2015rtl}%
  \BibitemOpen
  \bibfield  {author} {\bibinfo {author} {\bibfnamefont {B.}~\bibnamefont
  {Grinstein}}, \bibinfo {author} {\bibfnamefont {C.~W.}\ \bibnamefont
  {Murphy}},\ and\ \bibinfo {author} {\bibfnamefont {P.}~\bibnamefont
  {Uttayarat}},\ }\bibfield  {title} {\bibinfo {title} {{One-loop corrections
  to the perturbative unitarity bounds in the $CP$-conserving two-Higgs doublet
  model with a softly broken ${\mathbb{Z}}_2$ symmetry}},\ }\href
  {https://doi.org/10.1007/JHEP06(2016)070} {\bibfield  {journal} {\bibinfo
  {journal} {JHEP}\ }\textbf {\bibinfo {volume} {06}},\ \bibinfo {pages}
  {070}},\ \Eprint {https://arxiv.org/abs/1512.04567} {arXiv:1512.04567
  [hep-ph]} \BibitemShut {NoStop}%
\bibitem [{\citenamefont {Nie}\ and\ \citenamefont {Sher}(1999)}]{Nie:1998yn}%
  \BibitemOpen
  \bibfield  {author} {\bibinfo {author} {\bibfnamefont {S.}~\bibnamefont
  {Nie}}\ and\ \bibinfo {author} {\bibfnamefont {M.}~\bibnamefont {Sher}},\
  }\bibfield  {title} {\bibinfo {title} {{Vacuum stability bounds in the
  two-Higgs doublet model}},\ }\href
  {https://doi.org/10.1016/S0370-2693(99)00019-2} {\bibfield  {journal}
  {\bibinfo  {journal} {Phys. Lett. B}\ }\textbf {\bibinfo {volume} {449}},\
  \bibinfo {pages} {89} (\bibinfo {year} {1999})},\ \Eprint
  {https://arxiv.org/abs/hep-ph/9811234} {arXiv:hep-ph/9811234 [hep-ph]}
  \BibitemShut {NoStop}%
\bibitem [{\citenamefont {Akeroyd}\ \emph {et~al.}(2000)\citenamefont
  {Akeroyd}, \citenamefont {Arhrib},\ and\ \citenamefont
  {Naimi}}]{Akeroyd:2000wc}%
  \BibitemOpen
  \bibfield  {author} {\bibinfo {author} {\bibfnamefont {A.~G.}\ \bibnamefont
  {Akeroyd}}, \bibinfo {author} {\bibfnamefont {A.}~\bibnamefont {Arhrib}},\
  and\ \bibinfo {author} {\bibfnamefont {E.}~\bibnamefont {Naimi}},\ }\bibfield
   {title} {\bibinfo {title} {{Note on tree-level unitarity in the general two
  Higgs doublet model}},\ }\href
  {https://doi.org/10.1016/S0370-2693(00)00962-X} {\bibfield  {journal}
  {\bibinfo  {journal} {Phys. Lett. B}\ }\textbf {\bibinfo {volume} {490}},\
  \bibinfo {pages} {119} (\bibinfo {year} {2000})},\ \Eprint
  {https://arxiv.org/abs/hep-ph/0006035} {arXiv:hep-ph/0006035 [hep-ph]}
  \BibitemShut {NoStop}%
\bibitem [{\citenamefont {Mahmoudi}(2009)}]{Mahmoudi:2008tp}%
  \BibitemOpen
  \bibfield  {author} {\bibinfo {author} {\bibfnamefont {F.}~\bibnamefont
  {Mahmoudi}},\ }\bibfield  {title} {\bibinfo {title} {{SuperIso v2.3: A
  program for calculating flavor physics observables in supersymmetry}},\
  }\href {https://doi.org/10.1016/j.cpc.2009.02.017} {\bibfield  {journal}
  {\bibinfo  {journal} {Comput. Phys. Commun.}\ }\textbf {\bibinfo {volume}
  {180}},\ \bibinfo {pages} {1579} (\bibinfo {year} {2009})},\ \Eprint
  {https://arxiv.org/abs/0808.3144} {arXiv:0808.3144 [hep-ph]} \BibitemShut
  {NoStop}%
\bibitem [{\citenamefont {Bechtle}\ \emph {et~al.}(2014)\citenamefont
  {Bechtle}, \citenamefont {Heinemeyer}, \citenamefont {St{\r a}l},
  \citenamefont {Stefaniak},\ and\ \citenamefont {Weiglein}}]{Bechtle:2013xfa}%
  \BibitemOpen
  \bibfield  {author} {\bibinfo {author} {\bibfnamefont {P.}~\bibnamefont
  {Bechtle}}, \bibinfo {author} {\bibfnamefont {S.}~\bibnamefont {Heinemeyer}},
  \bibinfo {author} {\bibfnamefont {O.}~\bibnamefont {St{\r a}l}}, \bibinfo
  {author} {\bibfnamefont {T.}~\bibnamefont {Stefaniak}},\ and\ \bibinfo
  {author} {\bibfnamefont {G.}~\bibnamefont {Weiglein}},\ }\bibfield  {title}
  {\bibinfo {title} {{\texttt{HiggsSignals}: Confronting arbitrary Higgs
  sectors with measurements at the Tevatron and the LHC}},\ }\href
  {https://doi.org/10.1140/epjc/s10052-013-2711-4} {\bibfield  {journal}
  {\bibinfo  {journal} {Eur. Phys. J. C}\ }\textbf {\bibinfo {volume} {74}},\
  \bibinfo {pages} {2711} (\bibinfo {year} {2014})},\ \Eprint
  {https://arxiv.org/abs/1305.1933} {arXiv:1305.1933 [hep-ph]} \BibitemShut
  {NoStop}%
\bibitem [{\citenamefont {Bechtle}\ \emph {et~al.}(2020)\citenamefont
  {Bechtle}, \citenamefont {Dercks}, \citenamefont {Heinemeyer}, \citenamefont
  {Klingl}, \citenamefont {Stefaniak}, \citenamefont {Weiglein},\ and\
  \citenamefont {Wittbrodt}}]{Bechtle:2020pkv}%
  \BibitemOpen
  \bibfield  {author} {\bibinfo {author} {\bibfnamefont {P.}~\bibnamefont
  {Bechtle}}, \bibinfo {author} {\bibfnamefont {D.}~\bibnamefont {Dercks}},
  \bibinfo {author} {\bibfnamefont {S.}~\bibnamefont {Heinemeyer}}, \bibinfo
  {author} {\bibfnamefont {T.}~\bibnamefont {Klingl}}, \bibinfo {author}
  {\bibfnamefont {T.}~\bibnamefont {Stefaniak}}, \bibinfo {author}
  {\bibfnamefont {G.}~\bibnamefont {Weiglein}},\ and\ \bibinfo {author}
  {\bibfnamefont {J.}~\bibnamefont {Wittbrodt}},\ }\bibfield  {title} {\bibinfo
  {title} {{\texttt{HiggsBounds-5}: testing Higgs sectors in the LHC $13~
  \text{TeV}$ Era}},\ }\href {https://doi.org/10.1140/epjc/s10052-020-08557-9}
  {\bibfield  {journal} {\bibinfo  {journal} {Eur. Phys. J. C}\ }\textbf
  {\bibinfo {volume} {80}},\ \bibinfo {pages} {1211} (\bibinfo {year}
  {2020})},\ \Eprint {https://arxiv.org/abs/2006.06007} {arXiv:2006.06007
  [hep-ph]} \BibitemShut {NoStop}%
\bibitem [{\citenamefont {Hou}\ \emph {et~al.}(2019)\citenamefont {Hou},
  \citenamefont {Xie}, \citenamefont {Gao}, \citenamefont {Dulat},
  \citenamefont {Guzzi}, \citenamefont {Hobbs}, \citenamefont {Huston},
  \citenamefont {Nadolsky}, \citenamefont {Pumplin}, \citenamefont {Schmidt},
  \citenamefont {Sitiwaldi}, \citenamefont {Stump},\ and\ \citenamefont
  {Yuan}}]{Hou:2019qau}%
  \BibitemOpen
  \bibfield  {author} {\bibinfo {author} {\bibfnamefont {T.-J.}\ \bibnamefont
  {Hou}}, \bibinfo {author} {\bibfnamefont {K.}~\bibnamefont {Xie}}, \bibinfo
  {author} {\bibfnamefont {J.}~\bibnamefont {Gao}}, \bibinfo {author}
  {\bibfnamefont {S.}~\bibnamefont {Dulat}}, \bibinfo {author} {\bibfnamefont
  {M.}~\bibnamefont {Guzzi}}, \bibinfo {author} {\bibfnamefont {T.~J.}\
  \bibnamefont {Hobbs}}, \bibinfo {author} {\bibfnamefont {J.}~\bibnamefont
  {Huston}}, \bibinfo {author} {\bibfnamefont {P.}~\bibnamefont {Nadolsky}},
  \bibinfo {author} {\bibfnamefont {J.}~\bibnamefont {Pumplin}}, \bibinfo
  {author} {\bibfnamefont {C.}~\bibnamefont {Schmidt}}, \bibinfo {author}
  {\bibfnamefont {I.}~\bibnamefont {Sitiwaldi}}, \bibinfo {author}
  {\bibfnamefont {D.}~\bibnamefont {Stump}},\ and\ \bibinfo {author}
  {\bibfnamefont {C.-P.}\ \bibnamefont {Yuan}},\ }\bibfield  {title} {\bibinfo
  {title} {{Progress in the CTEQ-TEA NNLO global QCD analysis}},\ }\href@noop
  {} {\  (\bibinfo {year} {2019})},\ \Eprint {https://arxiv.org/abs/1908.11394}
  {arXiv:1908.11394 [hep-ph]} \BibitemShut {NoStop}%
\bibitem [{\citenamefont {Pumplin}\ \emph {et~al.}(2001)\citenamefont
  {Pumplin}, \citenamefont {Stump}, \citenamefont {Brock}, \citenamefont
  {Casey}, \citenamefont {Huston}, \citenamefont {Kalk}, \citenamefont {Lai},\
  and\ \citenamefont {Tung}}]{Pumplin:2001ct}%
  \BibitemOpen
  \bibfield  {author} {\bibinfo {author} {\bibfnamefont {J.}~\bibnamefont
  {Pumplin}}, \bibinfo {author} {\bibfnamefont {D.}~\bibnamefont {Stump}},
  \bibinfo {author} {\bibfnamefont {R.}~\bibnamefont {Brock}}, \bibinfo
  {author} {\bibfnamefont {D.}~\bibnamefont {Casey}}, \bibinfo {author}
  {\bibfnamefont {J.}~\bibnamefont {Huston}}, \bibinfo {author} {\bibfnamefont
  {J.}~\bibnamefont {Kalk}}, \bibinfo {author} {\bibfnamefont {H.~L.}\
  \bibnamefont {Lai}},\ and\ \bibinfo {author} {\bibfnamefont {W.~K.}\
  \bibnamefont {Tung}},\ }\bibfield  {title} {\bibinfo {title} {{Uncertainties
  of predictions from parton distribution functions. II. The Hessian method}},\
  }\href {https://doi.org/10.1103/PhysRevD.65.014013} {\bibfield  {journal}
  {\bibinfo  {journal} {Phys. Rev. D}\ }\textbf {\bibinfo {volume} {65}},\
  \bibinfo {pages} {014013} (\bibinfo {year} {2001})},\ \Eprint
  {https://arxiv.org/abs/hep-ph/0101032} {arXiv:hep-ph/0101032 [hep-ph]}
  \BibitemShut {NoStop}%
\end{thebibliography}%

\end{document}